\documentclass[aps,pra,twocolumn,showpacs,superscriptaddress]{revtex4}
\usepackage{graphicx}
\usepackage{amsmath}
\usepackage{amssymb}
\usepackage{lscape}
\usepackage{multirow}
\usepackage{longtable}

\input{epsf}

\begin{document}

\title{Three-boson spectrum in the 
presence of 1D spin-orbit coupling: Efimov's generalized
radial scaling law}

\author{Q. Guan}
\affiliation{Homer L. Dodge Department of Physics and Astronomy,
  The University of Oklahoma,
  440 W. Brooks Street,
  Norman,
Oklahoma 73019, USA}

\author{D. Blume}
\affiliation{Homer L. Dodge Department of Physics and Astronomy,
  The University of Oklahoma,
  440 W. Brooks Street,
  Norman,
Oklahoma 73019, USA}

\date{\today}

\begin{abstract}
Spin-orbit coupled cold atom systems, governed by Hamiltonians that
contain quadratic kinetic energy terms typical 
for a particle's motion in the usual Schr\"odinger equation and
linear kinetic energy terms typical
for a particle's motion in the usual Dirac equation,
have attracted a great deal of attention recently
since they provide an alternative route for realizing fractional
quantum Hall physics,
topological insulators, and spintronics physics.
The present work focuses on the three-boson system in the presence of
1D spin-orbit coupling,
which is most relevant
to ongoing cold atom experiments.
In the absence of spin-orbit coupling terms, the three-boson system
exibits the Efimov effect: the entire energy spectrum is uniquely
determined 
by the $s$-wave scattering length and a single three-body parameter, i.e.,
using one of the energy levels as input, the other energy levels
can be obtained via Efimov's radial scaling law, which is 
intimately tied to a
discrete scaling symmetry.
It is demonstrated that the discrete scaling symmetry persists in the
presence of 1D spin-orbit coupling, implying the
validity of a generalized radial scaling law in five-dimensional
space.
The dependence of the energy levels on the scattering length,
spin-orbit coupling parameters, and center-of-mass momentum
is discussed.
It is conjectured that three-body systems with other
types of spin-orbit coupling terms are also governed by generalized
radial scaling laws, provided the system exhibits the Efimov
effect in the absence of spin-orbit coupling.

\end{abstract}

\maketitle

\section{Introduction}
\label{sec1}
Under which conditions do two, three, or more particles 
form weakly-bound states, i.e., bound states
that are larger than the range of the two-, three-, and higher-body
forces that bind the particles together?
And under which conditions are the characteristics of these few-body
bound states governed by underlying symmetries?
These questions are of utmost importance across physics.
For example, 
the existence of bound tetra-quark systems~\cite{review_tetraquark}, first
proposed in 1964 by Gell-Mann~\cite{gell},
has been challenging our understanding of QCD.
The existence of the extremely weakly-bound triton
has a profound effect on the nuclear chart, including the 
existence of larger exotic halo nuclei~\cite{epelbaum_review,review_tanihata}.
Historically, the triton has played an important role
in the context of the Thomas collapse~\cite{thomas_collapse} and the
Efimov effect~\cite{efimov70,efimov71}, which is intimately 
tied to a discrete scaling symmetry of the three-body
Schr\"odinger equation.

The three-boson system with two-body short-range interactions
is considered the holy grail of few-body physics.
It has captured physicists' attention since Efimov's bizarre and 
counterintuitive
predictions in the early 70ies~\cite{efimov70,efimov71}
and has spurred a flurry of theoretical and experimental
works from nuclear to atomic to condensed matter to particle
physics~\cite{hammer2010, braaten2006, braaten2007, ferlainoPhysics,greenePhysicsToday, naidon2016, grimm2006, grimm2011, huangPRL, efimov_helium, efimov_radio1, efimov_radio2, Zaccanti, nishida}.
The unique scaling laws exhibited by Efimov trimers
can be traced back to the existence
of just one large length scale in the problem, namely
the two-body $s$-wave scattering length.
The main focus of the present work is on
investigating what happens to the three-boson Efimov states
in the presence of 1D spin-orbit coupling.
Similar to few-body systems on the lattice~\cite{mattis1986},
the 1D spin-orbit coupling introduces a
parametric dependence of the relative Hamiltonian
on the center-of-mass momentum.
This center-of-mass momentum dependence
leads, as we will show, to a modification
of the lowest break-up threshold and has a profound effect on the binding
energy.
Despite this dependence on the center-of-mass momentum and despite the
fact that the spin-orbit coupling terms depend on three 
additional parameters (namely, $k_{\text{so}}$, $\Omega$ and $\delta$; see below),
it is argued that the three-boson system in the presence of 
1D spin-orbit coupling possesses, in the zero-range limit,
a discrete scaling symmetry and it is shown that the energy spectrum 
is described by
a generalized radial scaling law. 

The 1D spin-orbit coupling terms,
which
break the rotational symmetry,
introduce
an unusual single-particle dispersion.
The Hamiltonian $\hat{H}_j$ 
of the $j$-th particle 
with mass $m$
and momentum operator $\hat{\vec{p}}_j$ 
(with components $\hat{p}_{j,x}$, $\hat{p}_{j,y}$, and $\hat{p}_{j,z}$)
is not simply given by  $\hat{\vec{p}}_j^2/(2m)$
but includes
a term that emulates a 
spin-1/2 particle interacting with a
momentum-dependent ``magnetic
field'' of infinite 
range~\cite{dalibard2011, galitski_review2013, review_hui, goldman2014, review_goldman2017},
\begin{eqnarray}
\label{eq_1dsoc}
\hat{H}_j=\frac{\hat{\vec{p}}_j^2}{2m} I_j + 
\hat{\vec{B}}(\hat{p}_{j,z}) \cdot \hat{\vec{\sigma}}_j.
\end{eqnarray}
Here, $I_j$
denotes the 2x2 identity matrix that spans the
spin degrees of freedom of the $j$-th particle,
the vector  $\hat{\vec{\sigma}}_j$ 
contains the three Pauli matrices $\hat{\sigma}_{j,x}$,
$\hat{\sigma}_{j,y}$, and $\hat{\sigma}_{j,z}$ of the $j$-th particle,
and $\hat{\vec{B}}$ 
represents
the effective
magnetic field, $\hat{\vec{B}}=(\Omega/2,0,\hbar k_{\text{so}} \hat{p}_{j,z}/m + \delta/2)$,
felt by the $j$-th particle.
The Raman coupling 
$\Omega$, detuning $\delta$, and 
spin-orbit coupling strength $k_{\text{so}}$,
which characterize the
two-photon Raman transition that couples (effectively)
two hyperfine states of an ultracold atom,
 describe the deviations from the 
``normal'' quadratic single-particle dispersion curves,
\begin{eqnarray}
E_{j,\pm}= \frac{\vec{p}_j^2}{2m} \pm \sqrt{
\left( \frac{\hbar k_{\text{so}} p_{j,z}}{m} + \frac{\delta}{2}\right)^2
+\frac{\Omega^2}{4}},
\end{eqnarray}
where $\vec{p}_j$ and  $p_{j,z}$ (both without ``hat'')
are expectation values of the corresponding 
operators.
For large $|\vec{p}_j|$, 
the dispersion curves 
$E_{j,\pm}$ approach $\vec{p}_j^2/(2m)$.
For small $|\vec{p}_j|$, in contrast,
the $E_{j,\pm}$ curves deviate appreciably
from $\vec{p}_j^2/(2m)$.
The momenta $\vec{p}_j$ are generalized
momenta (sometimes also referred to as quasi-momenta) and not mechanical
momenta (sometimes also referred to as kinetic 
momenta)~\cite{spielman}.
Throughout this article, we frequently drop the prefix 
``generalized''
and refer to $\vec{p}_j$ as momentum vector of the $j$-th atom.
The Hamiltonian given in Eq.~(\ref{eq_1dsoc}) can also
be realized by lattice shaking techniques as well as
in photonic crystals and mechanical 
setups~\cite{review_goldman2017, shakelattice1, shakelattice2, ozawa2018}.

If two-body short-range interactions are added, the modified single-particle
dispersion curves can 
significantly alter the properties of weakly-bound two-
and three-body states.
This has been demonstrated extensively for two identical 
fermions for 1D, 2D, and 3D spin-orbit
coupling~\cite{pairing_shenoy, rashbon_shenoy, bound_shenoy, crossover_shenoy, xiaoling, xiaoling2017, han_pu_pra, zhenhua, zhenhua2012, qingze}
and for two identical 
bosons for 2D and 3D spin-orbit coupling~\cite{wang2015, qingze, luo_boson, li_boson, xu_boson}
but not for the 1D spin-orbit coupling considered in this work.
The present work presents the first study  of
how the 
experimentally most frequently realized
1D spin-orbit coupling terms modify the 
three-boson energy spectrum.
We note, however, that several three-body studies
for bosonic and fermionic systems
with other types of spin-orbit coupling 
exist~\cite{cui2014,shi2014,shi2015,qiu2016}.
All of these earlier studies limited themselves to
vanishing center-of-mass momentum. Our work,
in contrast, allows for finite center-of-mass momenta.

The key objective of the present work is to show that the
three-boson system in the presence of 1D
spin-orbit coupling obeys a generalized radial scaling law, which reflects the
existence of a discrete scaling symmetry in the limit
of zero-range interactions.
The scaling parameter $\lambda_0$, $\lambda_0 \approx 22.694$,
is the same as in the absence of the spin-orbit coupling terms.
The generalized radial scaling law relates the energy for a 
given $1/a_s$, $k_{\text{so}}$, $\Omega$, and $\tilde{\delta}$
[$\tilde{\delta}$ is a generalized detuning that is defined in terms
of the detuning $\delta$ and 
the $z$-component of the center-of-mass momentum, 
see Eq.~(\ref{eq_deltatilde})]
to the energy for a scaled set of parameters, namely
$\lambda_0/a_s$, $\lambda_0 k_{\text{so}}$, $(\lambda_0)^2 \Omega$, and 
$(\lambda_0)^2 \tilde{\delta}$.
Correspondingly, the term ``radial'' does not refer to the radius 
in a two-dimensional space as in the usual
Efimov scenario but to the radius in a five-dimensional space.
The fact that the discrete scaling symmetry ``survives'' when the
spin-orbit coupling terms are added to the three-boson
Hamiltonian with zero-range interactions can be intuitively understood from
the observation that $k_{\text{so}}$, $\Omega$ and $\tilde{\delta}$
can be thought of as introducing finite length scales into the system.
In the standard Efimov scenario, $a_s$ introduces a finite length scale and the
radial scaling law holds regardless of whether $|a_s|$ is larger or smaller
than the size of the trimer, provided $|a_s|$ is much larger than the
intrinsic scales of the underlying two- and three-body interactions.
In the generalized Efimov scenario considered here,
the parameters $a_s$, $k_{\text{so}}$, $\Omega$, and $\tilde{\delta}$
each introduce a finite length scale. Correspondingly, the generalized radial
scaling law holds regardless of whether these length scales are larger or smaller than the
size of the trimer, provided the length scales are
much larger than the
intrinsic scales of the underlying two- and three-body interactions.

Our findings for the experimentally most frequently realized
1D spin-orbit coupling are 
consistent with 
Ref.~\cite{shi2015}.
References~\cite{shi2014,shi2015}
considered an impurity with 3D spin-orbit coupling
that interacts with two identical fermions that do not feel any spin-orbit
coupling
terms and interact with the impurity through 
short-range two-body potentials.
Restricting themselves to vanishing center-of-mass momenta, Ref.~\cite{shi2014}
stated that the trimers for mass ratio $\gtrsim 13.6$
``no longer obey the discrete scaling symmetry even at resonance''
because the spin-orbit coupling ``introduces an additional length scale''.
In Ref.~\cite{shi2015}, the 
same authors arrive at a 
seemingly
different conclusion, namely
``in the presence of SO [spin-orbit] coupling, the system exhibits 
a discrete scaling behavior''
and ``the scaling ratio is identical to that without SO [spin-orbit]
coupling''.
The two statements can be reconciled
by noting that the discrete scaling
symmetry requires an enlarged
parameter space, an aspect that was
recognized in Ref.~\cite{shi2015} but not in Ref.~\cite{shi2014}.
We conjecture that the discrete scaling symmetry holds for any type of
spin-obit coupling and all center-of-mass momenta.
Depending on the type of the spin-orbit coupling, the generalized
Efimov plot
is four- or five-dimensional and the generalized radial scaling law
applies to the entire low-energy spectrum.
The dependence of the energy levels on the system parameters has to be
calculated explicitly once for each type of spin-obit coupling.

The remainder of this article is organized as follows.
To set the stage, Sec.~\ref{sec_standardefimov}
reviews the standard Efimov scenario for three identical bosons.
Section~\ref{sec_symm_soc} introduces the 
system Hamiltonian in the presence of 1D spin-orbit coupling and
discusses the associated continuous and discrete scaling symmetries.
The generalized radial scaling law for the three-boson system 
in the presence of
1D spin-orbit coupling is confirmed numerically in Sec.~\ref{sec_numerics}.
Section~\ref{sec_dependence} 
highlights the role of the center-of-mass momentum and discusses 
possible experimental signatures of this dependence.
Finally,
Sec.~\ref{sec_conclusion} presents an outlook.
Technical details are relegated to several appendices.

\section{Review of standard Efimov scenario}
\label{sec_standardefimov}
The relative Hamiltonian
for two identical bosons of mass $m$ interacting through the
zero-range contact interaction
$V_{\text{2b,zr}}(\vec{r})$,
\begin{eqnarray}
  V_{\text{2b,zr}}(\vec{r}) = \frac{4 \pi \hbar^2 a_s}{m} \delta^{(3)}(\vec{r})
  \frac{\partial}{\partial r} r,
  \end{eqnarray}
where $a_s$ denotes the two-body $s$-wave scattering length
and $\vec{r}$ the internuclear distance vector
($r=|\vec{r}|$),
possesses a continuous scaling symmetry~\cite{braaten2006}.
Performing the
transformation 
\begin{align}
\label{eq_scale_c2}
a_s \rightarrow \lambda a_s, \;
\vec{r} \rightarrow \lambda \vec{r},
\mbox{ and }
t \rightarrow \lambda^2 t,
\end{align}
where 
$t$ denotes the time and $\lambda$ a real number (scaling parameter),
the relative two-body time-dependent Schr\"odinger equation
remains unchanged.

Importantly, the continuous scaling symmetry extends to three identical 
mass $m$ bosons
with position vectors $\vec{r}_j$ that interact
through pairwise $s$-wave zero-range interactions 
$V_{{\text{2b,zr}}}(\vec{r}_{jk})$~\cite{braaten2006}.
  To see this, we consider the time-dependent
  Schr\"odinger equation
  for
  the relative three-body Hamiltonian $\hat{H}_{\text{rel}}$,
  \begin{align}
\label{eq_3b_timedep}
    \hat{H}_{\text{rel}} = \sum_{j=1,2} -\frac{\hbar^2}{2 \mu_j} \nabla_{\vec{\rho}_j}^2
    +\sum_{j=1}^{2} \sum_{k=j+1}^3 V_{\text{2b,zr}} (\vec{r}_{jk})
    +V_{\text{3b,zr}}(R),
  \end{align}
  where $\vec{\rho}_{j}$ denotes the
  $j$-th relative Jacobi vector and $\mu_j$ the associated Jacobi mass.
  We use a ``K-tree'' (see Appendix~\ref{appendix_basics}) in which 
$\mu_1$ for the two-body system is given by 
$m/2$ and $\mu_1$ and $\mu_2$ for the three-body system are given by 
$m/2$ and $2m/3$.
  The zero-range three-body potential $V_{\text{3b,zr}}(R)$,
  \begin{align}
    V_{\text{3b,zr}}(R)=g_3 \frac{\hbar^2}{m} \delta^{(6)}(R),
  \end{align}
is written in terms
  of a six-dimensional delta-function in the three-body hyperradius $R$,
    $R^2 = r_{12}^2 + r_{13}^2 + r_{23}^2$.
  Since the coupling constant
  $g_3$ has units of $length^4$, it can be rewritten as
  $g_3 = C \kappa_*^{-4}$, where $C$ is a real
  constant and $\kappa_*$ the three-body
  binding momentum of one of the three-boson bound states at
  unitarity (infinite $a_s$).
  While $V_{\text{3b,zr}}(R)$ has to be regularized in practice,
  the explicit regularization is irrelevant for our purpose.
  Performing the transformation
\begin{align}
  \label{eq_scale_c3}
  a_s \rightarrow \lambda a_s, \;
  \vec{r}_{jk} \rightarrow \lambda \vec{r}_{jk}, \;
  t \rightarrow \lambda^2 t, 
  \mbox{ and } \kappa_* \rightarrow \lambda^{-1} \kappa_*,
  \end{align}
the Schr\"odinger equation for the
relative Hamiltonian given in
Eq.~(\ref{eq_3b_timedep}) remains unchanged, i.e., 
the three-body system possesses a continuous scaling symmetry.
  
\begin{figure}
\vspace*{0cm}
\hspace*{-0.8cm}
\includegraphics[width=0.4\textwidth]{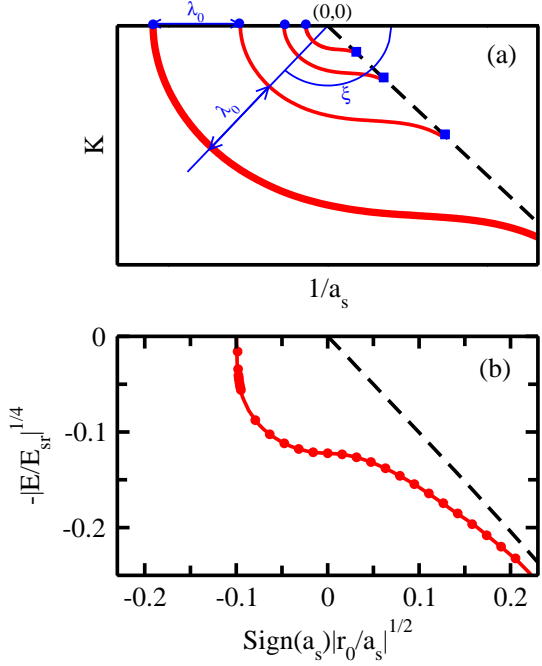}
\caption{(Color online)
Radial scaling law for the
standard Efimov scenario.
(a) The solid lines show the
quantity $K$ as a function of $1/a_s$ for the zero-range three-boson
Hamiltonian. To make this plot, $\lambda_0$ has been
artificially set to $2$ instead of $22.694$.
The dashed line shows the atom-dimer threshold.
The thin radially outgoing solid lines 
and arrows illustrate 
the scaling law.
Circles and squares mark the critical scattering
lengths $a_-$ at which the trimer energy
is degenerate with the three-atom threshold and 
the critical scattering lengths $a_*$ at which the
trimer energy is degenerate with the atom-dimer threshold, respectively.
(b) Collapse of neighboring energy levels for the
finite-range interaction model 
[$\hat{H}_{\text{rel}}$ in Eq.~(\ref{eq_3b_timedep})
with $V_{\text{2b,zr}}$ and $V_{\text{3b,zr}}$ replaced
by $V_{\text{2b,G}}$ and $V_{\text{3b,G}}$, respectively;
$R_0=\sqrt{8}r_0$ and $(\kappa_*)^{-1} \approx 66.05 r_0$].
The solid line shows the fourth-root
of the energy of the lowest three-boson
state as a function of the square-root of the
inverse of the $s$-wave scattering length. 
The dashed line shows the associated atom-dimer threshold.
The dots show the energy of the second-lowest
three-boson state, with the radial scaling law applied in 
reverse
so
as to collapse the second-lowest level
(dots) onto the lowest level (solid line).
For clarity, 
the scaled atom-dimer threshold 
for the second-lowest three-boson state is not shown.  
}
\label{fig_efimov_standard}
\end{figure}

Intriguingly, the three-body system with zero-range interactions
additionally exhibits
an exact discrete scaling symmetry~\cite{braaten2006}.
The discrete transformation is given by
\begin{align}
\label{eq_3b_scaled}
  a_s \rightarrow (\lambda_0)^n a_s, \;
  \vec{r}_{jk} \rightarrow (\lambda_0)^n \vec{r}_{jk}, \;
  t \rightarrow (\lambda_0)^{2n} t, \nonumber \\
  \mbox{ and } \kappa_* \rightarrow \kappa_*,
  \end{align}
where  
$n=\pm 1, \pm 2,\cdots,\pm \infty$ and
$\lambda_0 \approx 22.694$.
The discrete scaling transformation, which underlies the 
three-body Efimov effect, is illustrated in Fig.~\ref{fig_efimov_standard}(a).
Fixing the three-body parameter $\kappa_*$
[see Eq.~(\ref{eq_3b_scaled})],
the Efimov plot depicts $K$ as a function of $1/a_s$,
where 
\begin{align}
K = -\sqrt{m|E|/\hbar^2}
\end{align}
and $E$ denotes the eigen energy of the
Hamiltonian  $\hat{H}_{\text{rel}}$ given in
Eq.~(\ref{eq_3b_timedep}).
The thick solid line in Fig.~\ref{fig_efimov_standard}(a) 
shows $K$ for one of the three-body eigen
energies. 
The thick solid line merges
with the three-atom threshold on the negative $a_s$-side
and with the atom-dimer threshold (dashed line) on the positive
$a_s$-side.
The thick solid line is obtained by solving the time-independent
Schr\"odinger equation for the three-body Hamiltonian
$\hat{H}_{\text{rel}}$. 
Provided the thick solid line is known
(a parametrization can be found in Refs.~\cite{braaten2006, naidon2016}),
the thin solid lines---which correspond to other three-body
eigen energies---can be obtained using 
the discrete scaling symmetry
without having to explicitly solve the
Schr\"odinger equation again.
For the construction, it is convenient to switch from the vector $\vec{y}=(1/a_s,K)^T$
to a radius $y=|\vec{y}|$ and an angle $\xi$,
\begin{align}
\label{eq_radial1}
K = -y \sin \xi
\end{align}
and
\begin{align}
\label{eq_radial2}
(a_s)^{-1} = y \cos \xi,
\end{align}
where $\xi$ goes from $\pi/4$ to $\pi$.
The limits $\pi/4$ and $\pi$ are set by the
atom-dimer and three-atom thresholds, respectively.
To obtain the thin solid lines in Fig.~\ref{fig_efimov_standard}(a)
from the thick solid line, one fixes
the angle $\xi$ and reads off the values
of the pair $(1/a_s,K)$ corresponding to the 
solid line.
Using 
\begin{align}
y = \sqrt{ (a_s)^{-2} + K^2},
\end{align}
it can be seen that the discrete 
scaling transformation
$a_s \rightarrow (\lambda_0)^n a_s$ and 
$E \rightarrow (\lambda_0)^{-2n} E$ implies 
$y \rightarrow (\lambda_0)^{-n} y$. Thus, 
dividing the radius $y$ 
of the thick solid line 
by $(\lambda_0)^{\pm 1}, (\lambda_0)^{\pm 2},\cdots$
and using the scaled value of $y$
in Eqs.~(\ref{eq_radial1}) and (\ref{eq_radial2}),
one obtains the values of the vectors  
$\vec{y}=(1/a_s,K)^T$ corresponding to the thin solid lines. 
This construction, referred to as Efimov's radial
scaling law, is a direct consequence of the 
discrete scaling symmetry.
If the three-boson system is characterized by
$\kappa_*^{\text{new}}$ instead of $\kappa_*$, 
the entire energy spectrum is scaled, i.e.,
if $\vec{y}=(1/a_s,K)^T$ describes a point on the Efimov
plot for $\kappa_*$, then
$(\kappa_*^{\text{new}}/\kappa_*)\vec{y}$
describes a point on the Efimov plot for $\kappa_*^{\text{new}}$.

\section{Symmetries in the presence of 1D spin-orbit coupling}
\label{sec_symm_soc}
This section generalizes the
symmetry discussion presented in the previous section to
the two- and three-boson systems in the presence
of 1D spin-obit coupling.
As a first step, we derive the
relative two- and three-body Hamiltonian with zero-range interactions 
in the presence of 1D spin-orbit coupling.
In a second step,
it is shown that these systems possess a continuous scaling
symmetry.
In a third step, it is argued that the three-boson system
additionally exhibits a discrete scaling symmetry, suggesting the
existence of a generalized radial scaling law.
Numerical evidence that supports our claim that
the three-boson system with 1D spin-orbit coupling
is governed by a generalized radial scaling law is presented in
Sec.~\ref{sec_numerics}.

We start with the first step.
The $N$-boson Hamiltonian in the presence of
1D spin-orbit coupling reads
\begin{eqnarray}
\hat{H} =\hat{H}_{\text{ni}} + \hat{V}_{\text{int}},
\end{eqnarray}
where the non-interacting and interacting pieces are given by
\begin{eqnarray}
\hat{H}_{\text{ni}}&=& 
\sum_{j=1}^N 
\frac{\hat{\vec{p}}_j^2}{2m} I_{1,\cdots,N} + \nonumber \\
&&\sum_{j=1}^N \left( \frac{\hbar k_{\text{so}}}{m} \hat{p}_{j,z} + 
\frac{\delta}{2} \right) 
I_{1,\cdots,j-1}\hat{\sigma}_{j,z}I_{j+1,\cdots,N} + \nonumber \\ 
&& \sum_{j=1}^N \frac{\Omega}{2} I_{1,\cdots,j-1}
\hat{\sigma}_{j,x}I_{j+1,\cdots,N} 
\end{eqnarray}
and
\begin{align}
\label{eq_vint_soc}
  \hat{V}_{\text{int}}= \left( \sum_{j=1,j<k}^N V_{\text{2b}}(r_{jk}) 
  + \sum_{j=1,j<k<l}^N V_{\text{3b}}(r_{jkl}) \right) I_{1,\cdots,N}.
\end{align}
Here, $I_{j,\cdots,k}$ with $j<k$
spans the spin degrees of freedom
of particles $j$ through $k$,
$I_{j,\cdots,k}=I_j \otimes \cdots \otimes I_k$.
For $N=2$, only $V_{\text{2b}}$ contributes.
For $N=3$, $r_{jkl}$ is equal to the three-body hyperradius
$R$.
The interaction model considered throughout this work assumes that the
interactions are the same in all spin channels.

It is, just as in the case without spin-orbit coupling, convenient to 
use Jacobi coordinates
$\vec{\rho}_j$ and associated momentum operators $\hat{\vec{q}}_j$
instead of the single-particle quantities
$\vec{r}_j$ and $\hat{\vec{p}}_j$. 
Importantly, the $N$-th Jacobi ``quantities'' 
$\vec{\rho}_N$ and $\hat{\vec{q}}_N$
correspond to the center-of-mass vector and center-of-mass 
momentum
operator.
It can be shown straightforwardly that the
Hamiltonian $\hat{H}$ commutes with the center-of-mass
momentum operator $\hat{\vec{q}}_N$~\cite{qingze_thesis},
i.e., the Schr\"odinger equation
$\hat{H} \Psi = E \Psi$
can be solved for each fixed $\vec{q}_{N}$.
Using this and Jacobi coordinates, the non-interacting
fixed-$\vec{q}_N$ Hamiltonian,
denoted by $\hat{\bar{H}}_{\text{ni}}$,
reads
\begin{eqnarray}
\label{eq_hnijacobi1}
\hat{\bar{H}}_{\text{ni}}=\hat{\bar{H}}_{\text{ni},\text{rel}}+
\frac{\vec{q}_N^2}{2 \mu_N} I_{1,\cdots,N},
\end{eqnarray}
where
\begin{eqnarray}
\label{eq_hnijacobi2}
\hat{\bar{H}}_{\text{ni},\text{rel}}
&=& 
\sum_{j=1}^{N-1} 
\frac{\hat{\vec{q}}_j^2}{2 \mu_j} I_{1,\cdots,N} + \nonumber \\
&&\sum_{j=1}^{N-1}  \frac{\hbar k_{\text{so}}}{m} \hat{q}_{j,z} 
\hat{\Sigma}_{j,z} + \nonumber \\ 
&& \sum_{j=1}^N \frac{\Omega}{2} I_{1,\cdots,j-1}\hat{\sigma}_{j,x}
I_{j+1,\cdots,N} 
+ \nonumber \\
&& \left( \frac{\hbar k_{\text{so}}}{\mu_N} q_{N,z} + \frac{\delta}{2} \right)  \times
\nonumber \\
&&\left( \sum_{j=1}^N I_{1,\cdots,j-1} \otimes \hat{\sigma}_{j,z} \otimes 
I_{j+1,\cdots,N} \right).
\end{eqnarray}
The explicit form of the
operators $\hat{\Sigma}_{j,z}$ with $j=1,\cdots,N-1$
is given
in Appendix~\ref{appendix_basics}.
Note that the
first and second lines of 
Eq.~(\ref{eq_hnijacobi2}) contain momentum operators while
the fourth line of Eq.~(\ref{eq_hnijacobi2})
and
the second term on the right hand side of
Eq.~(\ref{eq_hnijacobi1})
contain expectation
values of the center-of-mass
momentum operators (and not operators).
As ``usual'', the interaction $\hat{V}_{\text{int}}$
depends on
$\vec{\rho}_1,\cdots,\vec{\rho}_{N-1}$ but not on the 
center-of-mass vector $\vec{\rho}_N$.
This implies that the eigen states
$\Psi$
can be written as
\begin{eqnarray}
  \Psi = \Phi_{\text{cm}} \Phi_{\text{rel}},
\end{eqnarray}
where~\cite{footnote2}
\begin{eqnarray}
  \Phi_{\text{cm}}=\exp\left( \frac{\imath \vec{q}_N \cdot \vec{\rho}_N }{\hbar}\right)
\end{eqnarray}
and where the
$\Phi_{\text{rel}}$, which are
eigen states of 
${\hat{\bar{H}}}_{\text{rel}}$, 
\begin{eqnarray}
\label{eq_hamrelbar}
\hat{\bar{H}}_{\text{rel}}=\hat{\bar{H}}_{\text{ni},\text{rel}}+\hat{V}_{\text{int}},
\end{eqnarray}
depend on the Jacobi vectors
$\vec{\rho}_1,\cdots,\vec{\rho}_{N-1}$ and the spin degrees of freedom.

Equation~(\ref{eq_hnijacobi2}) shows that
the eigen energies of
$\hat{H}$
depend on the 
generalized detuning
$\tilde{\delta}$,
\begin{eqnarray}
\label{eq_deltatilde}
\frac{\tilde{\delta}}{2} =
\frac{\hbar k_{\text{so}}}{\mu_N} q_{N,z} + \frac{\delta}{2},
\end{eqnarray}
i.e.,
$q_{N,z}$ and $\delta$ enter as a combination
and not as
independent parameters. 
This observation suggests that 
the center-of-mass momentum may
play a decisive role in determining the 
characteristics of the weakly-bound two- and three-body states
(see also Refs.~\cite{han_pu_pra,pairing_shenoy}).
The parametric dependence of the Hamiltonian 
$\hat{\bar{H}}_{\text{rel}}$
on the $z$-component $q_{N,z}$ of the center-of-mass 
momentum is a direct
consequence of the fact that the presence of the spin-orbit coupling
breaks the Galilean invariance~\cite{review_hui}.
One immediate consequence of the
broken Galilean invariance is that knowing the energy of an
eigen state with $q_{N,z}=0$ does not, in general, suffice for predicting the
energy of an
eigen state with $q_{N,z} \ne 0$.
Importantly, the eigen states $\Psi$
depend, in general, explicitly on $q_{N,z}$ and $\delta$ and not
just on $\tilde{\delta}$.

We are now ready to address the second step.
Parametrizing the two-body interactions $V_{\text{2b}}$ by the zero-range
potential $V_{\text{2b,zr}}$,
the $N=2$ relative Hamiltonian depends on
four parameters, namely $a_s$, $k_{\text{so}}$, $\Omega$, and $\tilde{\delta}$.
It can be readily checked that the
corresponding time-dependent Schr\"odinger equation is
invariant under the transformation
\begin{align}
  \label{eq_scalec_n2soc}
  a_s \rightarrow \lambda a_s, \;
  k_{\text{so}} \rightarrow \lambda^{-1} k_{\text{so}}, \;
  \Omega \rightarrow \lambda^{-2} \Omega, \;
  \tilde{\delta} \rightarrow \lambda^{-2} \tilde{\delta}, \; \nonumber \\
  \vec{r} \rightarrow \lambda \vec{r}, \;
  \mbox{ and }
  t \rightarrow \lambda^2 t,
\end{align}
i.e., the $N=2$ system with zero-range interactions
possesses a continuous scaling
symmetry.
The
continuous scaling symmetry extends to the three-boson system
with zero-range interactions
[$V_{\text{2b}}=V_{\text{2b,zr}}$ and 
$V_{\text{3b}}=V_{\text{3b,zr}}$ in Eq.~(\ref{eq_vint_soc})]
in the presence of spin-orbit coupling since the
corresponding time-dependent
$N=3$ 
Schr\"odinger equation
is invariant under the transformation
\begin{align}
  \label{eq_scalec_n3soc}
  a_s \rightarrow \lambda a_s, \;
  k_{\text{so}} \rightarrow \lambda^{-1} k_{\text{so}}, \;
  \Omega \rightarrow \lambda^{-2} \Omega, \;
  \tilde{\delta} \rightarrow \lambda^{-2} \tilde{\delta}, \; \nonumber \\
  \vec{r} \rightarrow \lambda \vec{r}, \;
  t \rightarrow \lambda^2 t,
  \mbox{ and }
\kappa_* \rightarrow \lambda^{-1} \kappa_*.
\end{align}
Equations~(\ref{eq_scalec_n2soc})
and
(\ref{eq_scalec_n3soc})
generalize
Eqs.~(\ref{eq_scale_c2})
and
(\ref{eq_scale_c3}) from Sec.~\ref{sec_standardefimov}.

Paralleling the discussion of  Sec.~\ref{sec_standardefimov},
step three poses the question whether or not the
three-boson system in the presence of spin-orbit coupling additionally
possesses a discrete scaling symmetry in the zero-range interaction limit.
Our claim is that it does and that the discrete
transformation is given by
\begin{align}
  \label{eq_scaled_n3soc}
  a_s \rightarrow (\lambda_0)^n a_s, \;
  k_{\text{so}} \rightarrow (\lambda_0)^{-n} k_{\text{so}}, \;
  \Omega \rightarrow (\lambda_0)^{-2n} \Omega, \; \nonumber \\
  \tilde{\delta} \rightarrow (\lambda_0)^{-2n} \tilde{\delta}, \; 
  \vec{r} \rightarrow (\lambda_0)^n \vec{r}, \;
  t \rightarrow (\lambda_0)^{2n} t,
  \mbox{ and }
\kappa_* \rightarrow  \kappa_*,
\end{align}
where $\lambda_0$ is identical to the scaling factor
of the standard Efimov scenario, i.e., $\lambda_0 \approx 22.694$.
Since no general analytical solutions exist
to the three-boson Schr\"odinger equation in the presence
of spin-orbit coupling, we rely on numerics to
support our claim.
The claim that the discrete scaling symmetry
survives in the presence of the
spin-orbit coupling terms can be understood intuitively by
realizing that
the spin-orbit coupling terms modify the low- but not
the high-energy portions of the single-particle dispersion
curves.
To set the stage for the numerical calculations
presented in the next section,
we discuss a number of consequences
of the
discrete scaling symmetry.

The discrete scaling symmetry suggests a generalized
radial scaling law 
for the three-boson system in the presence
of 1D spin-orbit coupling 
in which the Efimov plot for $\vec{y}=(1/a_s,K)^T$ 
discussed in the previous section is replaced 
by a generalized Efimov plot
for
\begin{align}
\label{eq_5param}
\vec{y}=(1/a_s, k_{\text{so}},
\mbox{Sign}(\Omega)\sqrt{m |\Omega|/\hbar^2},
\mbox{Sign}(\tilde{\delta})\sqrt{m |\tilde{\delta}|/\hbar^2},
K)^T.
\end{align}
In the limit that the second, third, and fourth 
parameters vanish, each of the usual Efimov energies is
four-fold degenerate
due to the fact that the spin degrees of freedom enlarge the
three-boson Hilbert space by a factor of four 
(from the $2^3=8$ independent spin configurations,
one can construct four fully symmetric spin functions).
For non-vanishing $k_{\text{so}}$, $\Omega$, and $\tilde{\delta}$,
we expect that the three-boson system supports four ``unique''
energy levels.
Each of the four energies,
collectively referred to as a manifold,
is characterized by a vector $\vec{y}$.
Knowing the dependence of each of these energy curves on
$1/a_s$, 
$k_{\text{so}}$,
$\mbox{Sign}(\Omega)\sqrt{m |\Omega|/\hbar^2}$, and
$\mbox{Sign}(\tilde{\delta})\sqrt{m |\tilde{\delta}|/\hbar^2}$,
there should exist other energy manifolds for the same $\kappa_*$
that can be obtained from the manifold that has been mapped out
without explicitly solving the three-boson Schr\"odinger equation
again.

To see how, we switch 
from
the five parameters given in Eq.~(\ref{eq_5param})
to the length $y=|\vec{y}|$ and four angles
$\xi_1,\cdots,\xi_4$ for each of the four energy levels
in the ``reference manifold'',
\begin{align}
\label{eq_radialgen1}
K = & -y \sin \xi_1 \sin \xi_2 \sin \xi_3 \sin \xi_4, \\
\mbox{Sign}(\tilde{\delta})\sqrt{m|\tilde{\delta}| / \hbar^2} = &
y \cos \xi_1 \sin \xi_2 \sin \xi_3 \sin \xi_4,\\
\mbox{Sign}(\Omega)\sqrt{m|\Omega| / \hbar^2} = &
y \cos \xi_2 \sin \xi_3 \sin \xi_4,\\
k_{\text{so}}=&
y \cos  \xi_3 \sin \xi_4,
\end{align}
and
\begin{align}
\label{eq_radialgen5}
1/a_s = & y \cos  \xi_4.
\end{align}
The full range of possible $a_s$, $k_{\text{so}}$, $\Omega$,
and $\tilde{\delta}$ is covered if
$\xi_1,\xi_2,\xi_3,\xi_4 \in [0,\pi]$.
The range of the angles is further constrained by
the energy surfaces of the three-atom and atom-dimer
thresholds (see Secs.~\ref{sec_numerics} and \ref{sec_dependence}).
To obtain the $K$ for other manifolds,
one chooses a direction of the
vector $\vec{y}$ by fixing the angles $\xi_1$ to $\xi_4$
and reads off the values of the components of $\vec{y}$
for each of the four known energy levels.
Using
\begin{align}
y=\sqrt{
(a_s)^{-2}+
(k_{\text{so}})^2 + 
\frac{m |\Omega|}{\hbar^2} + 
\frac{m |\tilde{\delta}|}{\hbar^2}+
K^2
},
\end{align}
it can be seen that the discrete transformation
$a_s \rightarrow (\lambda_0)^n a_s$,
$k_{\text{so}} \rightarrow (\lambda_0)^{-n} k_{\text{so}}$,
$\Omega \rightarrow (\lambda_0)^{-2n} \Omega$,
$\tilde{\delta} \rightarrow (\lambda_0)^{-2n} \tilde{\delta}$,
$E \rightarrow (\lambda_0)^{-2n} E$
implies
$y \rightarrow (\lambda_0)^{-n} y$.
Thus, dividing the ``hyperradius'' $y$ 
corresponding to the $j$-th energy in the reference
manifold by $(\lambda_0)^{\pm 1}$, $(\lambda_0)^{\pm 2},\cdots$
and using the scaled value of $y$ in 
Eqs.~(\ref{eq_radialgen1})-(\ref{eq_radialgen5}),
one obtains the values of the components of
$\vec{y}$
for the $j$-th energy level in the other manifolds.
The generalized scaling law is tested in the next
section by considering two
neighboring energy manifolds and confirming that the energy manifolds collapse
onto each other if the discrete
scaling transformation is applied to
the energy levels in the more weakly-bound manifold.

\section{Numerical test of the generalized radial scaling law}
\label{sec_numerics}

To facilitate the numerical calculations, we replace the
two-body zero-range potential $V_{\text{2b,zr}}$ 
by
an attractive Gaussian $V_{\text{2b,G}}$
with range $r_0$ and depth $v_0$,
\begin{eqnarray}
V_{\text{2b,G}}(r_{jk}) = v_0 \exp \left( -\frac{r_{jk}^2}{2r_0^2}
\right),
\end{eqnarray}
where $v_0$ is negative and adjusted such 
that 
$V_{\text{2b,G}}(r_{jk})$ supports at most one two-body
$s$-wave bound state.
To reduce finite-range
effects, we aim to work in
the regime where the absolute value of the free-space
$s$-wave scattering length $a_s$
is notably larger than $r_0$.
Parameter combinations where the
absolute value
of the free-space 
$p$-wave 
scattering volume is
large are excluded. 
The three-body zero-range 
potential $V_{\text{3b,zr}}$ 
is replaced by a repulsive Gaussian $V_{\text{3b,G}}$
with range
$R_0$ and height $V_0$,
\begin{eqnarray}
V_{\text{3b,G}}(r_{jkl}) = V_0 \exp \left( -\frac{r_{jkl}^2}{2R_0^2}
\right).
\end{eqnarray}
In our numerical calculations, $R_0$ is fixed 
at $\sqrt{8}r_0$ and  $V_0$ ($V_0 \ge 0$) is varied to dial in the
desired three-body parameter $\kappa_*$.
Specifically, we define $\kappa_*$ to be the binding momentum
of the energetically lowest-lying universal three-body state
at unitarity (infinite $a_s$)  
for $k_{\text{so}}=\Omega=\tilde{\delta}=0$.
Without the repulsive three-body potential, the lowest three-body state
is not universal~\cite{yan2015}.
The repulsive three-body potential
pushes the lowest three-body energy up
and we adjust $V_0$, for fixed $v_0$ (infinite $a_s$),
such that the energy of the
lowest three-body state for finite $V_0$
is identical to the energy of the first excited three-body
state for $V_0=0$.
This corresponds to $(\kappa_*)^{-1} \approx 66.05r_0$, i.e.,
the trimer is much larger than the intrinsic scales of the two- and three-body
interactions.
With the repulsive three-body potential turned on,
the radial scaling law can be tested using
the two lowest-lying energy manifolds.

We start the discussion of our numerical results
by looking at the three-body spectrum 
for the standard Efimov scenario
($k_{\text{so}}=\Omega=\tilde{\delta}=0$).
The reason for discussing this ``reference system''
is two-fold: 
it
illustrates how to check the 
validity of the radial scaling law 
for a case where it is known to hold
and
it gives us a sense 
for the finite-range corrections expected in the
presence of spin-orbit coupling.
The solid line in Fig.~\ref{fig_efimov_standard}(b)
shows the relative three-body energy of the
energetically lowest-lying state
as a function of the inverse of the $s$-wave scattering length
for the Hamiltonian given in Eq.~(\ref{eq_3b_timedep}) with
$V_{\text{2b,zr}}$ and $V_{\text{3b,zr}}$ replaced by
$V_{\text{2b,G}}$ and $V_{\text{3b,G}}$, respectively.
To ``compress'' the data, the horizontal and vertical
axis employ a square-root and fourth-root representation.
The scattering length is scaled by $r_0$ and the energy by
$E_{\text{sr}}$,
\begin{align}
E_{\text{sr}}=\frac{\hbar^2}{m r_0^2}.
\end{align}
The trimer energy merges with the three-atom threshold
on the negative scattering length side at
$r_0/|a_s| \approx 0.01$.

To get a feeling for the finite-range effects, 
we assume that the
radial scaling law holds and apply it ``in reverse''.
Specifically, using numerically determined
pairs $(1/a_s, K)$ corresponding to the excited state,
the dots in Fig.~\ref{fig_efimov_standard}(b) show 
the points $(\lambda_0 r_0/a_s, \lambda_0 r_0 K)$,
using---as for the lowest state---the
square-root and fourth-root depiction.
In the zero-range limit
($r_0 \rightarrow 0$ and $R_0 \rightarrow 0$),
the dots would lie on top of the
solid line. 
The nearly
perfect agreement between the solid line and the dots
in Fig.~\ref{fig_efimov_standard}(b)
indicates that the finite-range effects 
are negligibly small for
the parameter combinations considered.

To test the generalized radial scaling law
proposed in Sec.~\ref{sec_symm_soc}, 
we calculate the eigen energies of states
in the lowest and second-lowest manifolds of
$\hat{\bar{H}}_{\text{rel}}$ 
(there are at most four states in each manifold)
and scale the energies in the second-lowest manifold
assuming that the generalized radial scaling law holds.
If the energy curves collapse,
the generalized radial scaling law is
validated.

\begin{figure}
\vspace*{0cm}
\hspace*{0.0cm}
\includegraphics[width=0.4\textwidth]{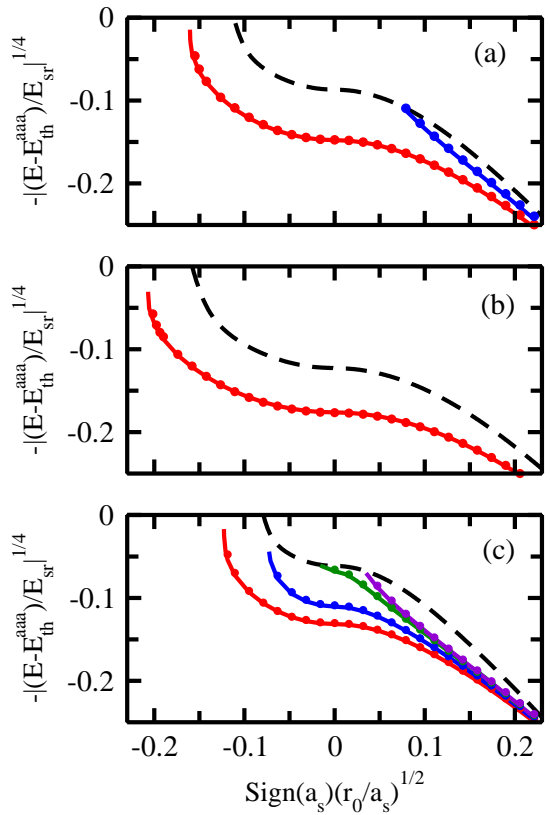}
\caption{(Color online)
Testing the generalized radial scaling law in the presence of 
1D spin-orbit coupling.
Panels~(a)-(c)
demonstrate the collapse of two neighboring energy manifolds for the
finite-range interaction model 
[$\hat{\bar{H}}_{\text{rel}}$ in Eq.~(\ref{eq_hamrelbar})
with $V_{\text{2b}}=V_{\text{2b,G}}$ and $V_{\text{3b}}=V_{\text{3b,G}}$;
$R_0=\sqrt{8}r_0$ and $(\kappa_*)^{-1} \approx 66.05 r_0$].
The energies of states in the lowest manifold (solid lines)
are obtained for
$\Omega= 2 E_{\text{so}}$, $\tilde{\delta}=0$, and
(a) $(k_{\text{so}})^{-1} = 50 r_0$, 
(b) $(k_{\text{so}})^{-1} = 25 r_0$, and 
(c) $(k_{\text{so}})^{-1} = 100 r_0$.
In all three panels, 
the dashed lines show the atom-dimer threshold.
The dots show the energies of states in the second-lowest
manifold, with the generalized radial scaling law applied in reverse so
as to collapse the 
three-body energies of states in the second-lowest manifold
(dots) onto the 
three-body 
energies of the states in the lowest manifold (solid lines).
For clarity, 
the scaled atom-dimer thresholds 
for the second-lowest energy manifold are not shown
in any of the panels.  
}
\label{fig_efimov_soc}
\end{figure}

In the presence of the 1D spin-orbit coupling,
the generalized Efimov plot has five axes.
Clearly, visualizing energy surfaces that depend on four parameters
is impossible and fully mapping out these high-dimensional 
dependences
is computationally demanding.
Thus, we consider selected cuts in the five-dimensional 
space.
Our first cut
uses $(k_{\text{so}})^{-1}=50r_0$,
$\Omega=2 E_{\text{so}}=0.04E_{\text{sr}}$,
where
\begin{align}
  E_{\text{so}} = \frac{(\hbar k_{\text{so}})^2}{2m},
\end{align}
and $\tilde{\delta}=0$.
For these parameters, we calculate the relative energy
$E$ of the states in the lowest energy manifold.
The solid lines in Fig.~\ref{fig_efimov_soc}(a)
show the quantity
$-|(E-E^{\text{aaa}}_{\text{th}})/E_{\text{sr}}|^{1/4}$ 
as a function
of $\mbox{Sign}(a_s)|r_0/a_s|^{1/2}$,
where $E^{\text{aaa}}_{\text{th}}$
denotes the energy of the lowest three-atom threshold
whose wave function has the same total momentum $q_{3,z}$ along
the $z$-axis
as the three-body system.
The determination of $E^{\text{aaa}}_{\text{th}}$ is discussed in
Appendix~\ref{appendix_threeatomthreshold}.
The energy $E^{\text{aaa}}_{\text{th}}$ is independent of $\kappa_*$ and
referencing $E$ relative to the lowest three-atom threshold
does not alter the generalized radial scaling law.
Figure~\ref{fig_efimov_soc}(a) shows that the lowest energy manifold
consists of, depending on the value of
$r_0/a_s$, zero, one, or two energy levels
[the second and third excited states of the lowest manifold exist at larger $r_0/a_s$
than those shown in Fig.~\ref{fig_efimov_soc}(a)].

The lowest three-body energy merges with the
three-atom threshold on the negative
$s$-wave scattering length side
and with the atom-dimer threshold
[dashed line in Fig.~\ref{fig_efimov_soc}(a)] 
on the positive scattering length side.
The determination of the atom-dimer threshold 
energy $E^{\text{ad}}_{\text{th}}$ is discussed
in Appendix~\ref{appendix_atomdimerthreshold}.
Just as the three-atom threshold, the atom-dimer threshold is
independent of $\kappa_*$ and determined such that
the momentum $q_{3,z}$ of the atom-dimer system is the same as that of the
three-body system.
The second lowest state
does not merge with the three-atom threshold on the negative $a_s$ side
but with the
atom-dimer threshold on the positive $a_s$ side.

Having determined the energies of the states in the lowest
energy manifold, the next step is to calculate the energies
of the states in the second-lowest energy manifold.
To map the energies
of the states in the second-lowest manifold onto the energies 
of the states in the lowest
energy manifold, we use the same $r_0$, $R_0$, and $\kappa_*$
and calculate the energies of the states
in the second-lowest energy
manifold for a $k_{\text{so}}$ that is $\lambda_0$
times smaller than the $k_{\text{so}}$ used to calculate the energies
of the states in the lowest energy manifold
[i.e., for $(k_{\text{so}})^{-1}\approx 1,135 r_0$],
for a $\Omega$ that is $(\lambda_0)^2$ times smaller than
the $\Omega$ used to calculate the energies
in the lowest energy manifold
(i.e., for $\Omega \approx 7.77 \times 10^{-7} E_{\text{sr}}$),
and
for $\tilde{\delta}=0$ (the scaling does not change zero)
as a function of $r_0/a_s$.
Having calculated
the
energies of the states in the second-lowest manifold
for
the scaled $k_{\text{so}}$,
$\Omega$, and $\tilde{\delta}$, 
the pairs
$(1/a_s,E-E^{\text{aaa}}_{\text{th}})$ 
are scaled
(note that $E^{\text{aaa}}_{\text{th}}$ for the excited state
manifold is calculated
using the scaled $k_{\text{so}}$, $\Omega$, and $\tilde{\delta}$
values).
The dots in Fig.~\ref{fig_efimov_soc}(a) show the
scaled pairs
$(\lambda_0 r_0/a_s,-(\lambda_0)^2 |E-E^{\text{aaa}}_{\text{th}}|/E_{\text{sr}})$,
  using---as for the lowest energy manifold---the square-root
  and fourth-root depiction.
  It can be seen that the solid lines and dots agree 
very well.
Note that the atom-dimer threshold for the second-lowest energy manifold
also needs to be recalculated using the scaled
$k_{\text{so}}$, $\Omega$, and $\tilde{\delta}$ 
[the resulting energies lie essentially on top
of the dashed line and are not shown in Fig.~\ref{fig_efimov_soc}(a)]. 
The deviation between the solid line and dots is $0.025$~\% for
$(r_0/a_s)^{1/2}=0$ and $0.80$~\% for $(r_0/a_s)^{1/2}=0.19$.
These deviations are
comparable to those between the corresponding atom-dimer thresholds
[in this case, the deviations are
$0.023$~\% for
$(r_0/a_s)^{1/2}=0$ and $0.82$~\% for $(r_0/a_s)^{1/2}=0.19$].
We conclude that our numerical results are consistent with 
the generalized radial scaling law.

  We emphasize that the scaling law has to be applied to all five axes
  of the generalized Efimov plot, i.e., to obtain the dots
in Fig.~\ref{fig_efimov_soc}(a)
  it is imperative to not only scale the two axes depicted but
  also the parameters corresponding
  to the three axes that are not depicted.
The ratio of the lowest energy in neighboring manifolds
at unitarity, e.g., is only equal to $22.694^2$ if 
the direction of $\hat{y}$ is the same for the two
energy levels under consideration.

The energy scales $E_{\text{so}}$, $|\Omega|$, and
$|\tilde{\delta}|$ are much smaller than
$|E-E^{\text{aaa}}_{\text{th}}|$ for a large portion of 
Fig.~\ref{fig_efimov_soc}(a). 
The region close to the
three-atom threshold is an exception. 
As such it might be argued that the
spin-orbit coupling terms are too weak to notably influence the
energy spectrum, possibly suggesting that the applicability of the
generalized radial scaling law is trivial.
One fact that speaks against this argumentation is that the
shape of the energy levels is notably influenced by the 
spin-orbit coupling terms. This is, e.g., reflected by the fact that
the energy levels in a given energy manifold are not degenerate.
To more explicitly demonstrate that the generalized radial scaling law 
holds when one or more of the energy scales 
associated with the spin-orbit coupling terms is/are larger than the 
binding energy, we repeat the calculations
for larger $k_{\text{so}}$ than those used in Fig.~\ref{fig_efimov_soc}(a).
Specifically, to determine the energy of the lowest state
in the lowest energy manifold
[solid line in Fig.~\ref{fig_efimov_soc}(b)], 
we use $(k_{\text{so}})^{-1}=25 r_0$
while keeping $r_0$, $R_0$, $\kappa_*$, and $\tilde{\delta}$
unchanged. The Raman coupling strength $\Omega$ is 
set to be equal to
$2 E_{\text{so}}$.
To demonstrate the collapse of 
the energies of the lowest states in the second-lowest 
and lowest manifolds, 
we apply the generalized
radial scaling law in the same way as in 
Fig.~\ref{fig_efimov_soc}(a).
The energy of the lowest state
in the second-lowest manifold 
is shown by dots in Fig.~\ref{fig_efimov_soc}(b). 
The agreement with the solid
line is excellent, 
supporting our claim that the generalized radial scaling law is not
limited to the case where the energy scales associated with the
spin-orbit coupling are smaller than the binding energy of the trimer,
provided these energies are notably smaller than $E_{\text{sr}}$.

To show the characteristics of the excited states
in the lowest manifold
in more detail, we consider a smaller $k_{\text{so}}$,
$(k_{\text{so}})^{-1}=100 r_0$, and as before
$\tilde{\delta}=0$ and $\Omega=2E_{\text{so}}$.
The use of a smaller $k_{\text{so}}$ [solid lines
in Fig.~\ref{fig_efimov_soc}(c)] moves the merging points of the
three-body energies corresponding to the excited
states
with the atom-dimer threshold to the
left compared to Fig.~\ref{fig_efimov_soc}(a).
Again, scaling the parameters appropriately, the
dots in Fig.~\ref{fig_efimov_soc}(c) 
show the energies of the states in the
second-lowest energy manifold.
The dots agree nearly perfectly with the
solid lines not only for the lowest
state in the two manifolds but also for the excited states in the 
two manifolds,
lending strong numerical support for the validity of the
generalized radial scaling law and hence for the
existence of the discrete scaling symmetry 
in the presence of 1D spin-orbit coupling terms
in the zero-range limit.

As already mentioned, the three-body parameter $\kappa_*$ for $V_0=0$,
defined using the energy of the first excited state
at unitarity in the absence of spin-orbit coupling, is identical to the
$\kappa_*$ for the 
three-body interaction 
with finite $V_0$ 
used throughout this section.
Turning on the spin-orbit coupling, we checked that the energies
of states in the second-lowest manifold for $V_0=0$ agree well with the
energies of states in the lowest manifold for the finite $V_0$.
This provides evidence that the generalized radial scaling law
is, just as the standard radial scaling law, independent
of the details of the underlying microscopic interaction model.
To confirm the continuous scaling symmetry of the
three-body Hamiltonian in the presence of
1D spin-orbit coupling, we checked 
that the energies for different $\kappa_*$ can be mapped onto
each other:
If $\vec{y}$ describes a point on the Efimov plot for $\kappa_*$,
then $(\kappa_*^{\text{new}}/\kappa_*) \vec{y}$ describes a point
on the Efimov plot for the new $\kappa_*^{\text{new}}$.

\section{Experimental implications: Role of center-of-mass momentum}
\label{sec_dependence}
Measuring signatures associated with two consecutive 
trimer energy levels is challenging, especially for 
equal-mass bosons, due to the relatively large discrete
scaling factor of $22.694$. The reason is 
that the absolute value
of the scattering length should, on the one hand, 
be notably larger than the van der Waals 
length $r_{\text{vdW}}$ and, on the other hand,
be smaller than the de Broglie
wave length $\lambda_{\text{dB}}$~\cite{braaten2006,naidon2016}. 
Despite these challenges,
the discrete 
scaling symmetry underlying the standard Efimov scenario has been confirmed 
experimentally by monitoring the atom losses
of an ultracold thermal gas of Cs atoms as a function
of the $s$-wave scattering length~\cite{huangPRL}
(for unequal mass mixtures, see Refs.~\cite{weidemueller,chin}). 
When the trimer energy is degenerate
with the three-atom threshold
[dots in Fig.~\ref{fig_efimov_standard}(a); the corresponding
critical scattering lengths are denoted by $a_{-}^{(n)}$]
or with the atom-dimer threshold
[squares in Fig.~\ref{fig_efimov_standard}(a); the corresponding
critical scattering lengths are denoted by $a_{*}^{(n)}$],
the losses are enhanced.
Since the critical scattering lengths for consecutive
trimer states are related to the scaling factor $\lambda_0$,
these atom-loss measurements provide 
a direct confirmation of the discrete scaling symmetry.
In addition, other characteristics of the standard Efimov scenario have been
measured~\cite{braaten2006, naidon2016,ferlainoPhysics,greenePhysicsToday}. 
For example,
the critical scattering lengths $a_{-}^{(n)}$ and $a_{*}^{(n)}$ for a given
trimer level $n$ are related to each other
by a universal number.
Correspondingly, the experimentally determined ratio 
$a_{-}^{(n)}/a_{*}^{(n)}$ can be viewed as a test of the functional form
of the energy levels shown in Fig.~\ref{fig_efimov_standard}(a).
Other experimental tests of the standard Efimov scenario include
the determination of the binding energy of an Efimov trimer via radio-frequency
spectroscopy~\cite{efimov_radio1, efimov_radio2}, 
the imaging of the quantum
mechanical density of the helium Efimov trimer via Coulomb explosion~\cite{efimov_helium},
and the observation of four- and five-body loss features that are
universally linked to the critical scattering lengths 
of the Efimov trimer~\cite{ferlaino2008,four_body, pollack2009,five_body}.

Directly measuring the discrete scaling symmetry
in the presence of spin-orbit coupling requires varying 
the inverse of the
$s$-wave scattering length by the scaling factor
$\lambda_0$, as in the standard Efimov
scenario, as well as 
varying the spin-orbit coupling parameters
$\Omega$, $E_{\text{so}}$, and $\tilde{\delta}$ by $(\lambda_0)^2$.
Covering such a wide range of
parameters is expected to be very challenging experimentally.
In what follows we instead focus on the situation where
the spin-orbit coupling parameters $k_{\text{so}}$ and $\Omega$
are held fixed while the $s$-wave scattering length $a_s$
and generalized detuning $\tilde{\delta}$ are varied.
An analogous study for the standard Efimov scenario would
look at the three-boson system for
a fixed finite $s$-wave scattering length. In this case, the
energy spacing would not be $(\lambda_0)^2$; however, the energies
of neighboring states would still be uniquely
related to each other.

For concreteness, we consider the $^{133}$Cs
system~\cite{footnote}, 
for which the three-atom resonances 
in the absence of spin-orbit coupling
occur at the
critical scattering lengths 
$a_-^{(0)} \approx -936 a_0$ and $a_-^{(1)} \approx -20,190a_0$~\cite{huangPRL}.
Here, $a_0$ denotes the Bohr
radius and the superscripts ``$(0)$'' and ``$(1)$'' indicate that these
critical scattering lengths are for the ground 
and first excited Efimov trimers, respectively.
Applying our numerical result $\kappa_* a_- = -1.505$
to the first excited state, the Cs system is characterized by
$(\kappa_*)^{-1} \approx 13,416 a_0$.

\begin{figure}
\vspace*{0.5cm}
\hspace*{-0.5cm}
\includegraphics[width=0.32\textwidth]{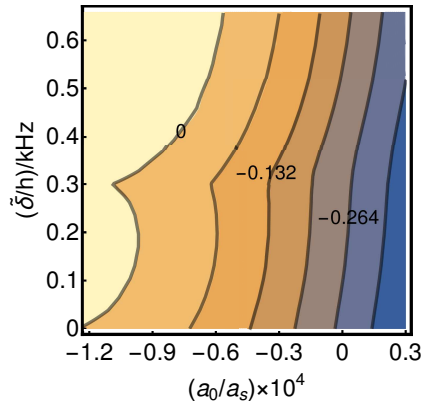}
\caption{(Color online)
The contours show the negative
of the three-boson binding energy, in kHz, of the lowest
state in the second lowest manifold
  as functions of $(a_s)^{-1}$ and $\tilde{\delta}$
for $k_{\text{so}}/\kappa_* \approx 1.32$ and $\Omega = 2 E_{\text{so}}$, where
$\kappa_*$ denotes the binding momentum of the first
excited Efimov trimer at unitarity in the
absence of spin-orbit coupling.
The conversion to $a_0$ and kHz is done using the
experimentally determined 
value of $a_-^{(1)}$ for Cs (see text).
The calculations are performed for
$\hat{\bar{H}}_{\text{rel}}$ with $V_{\text{2b}}=V_{\text{2b,G}}$ and $V_{\text{3b}}=0$
  [$(\kappa_*)^{-1} \approx 66.05 r_0$].
}
\label{fig_threebody_energy}
\end{figure}

Figure~\ref{fig_threebody_energy}
shows the negative of the binding energy of the lowest state in the 
first excited manifold
for $k_{\text{so}}/\kappa_* \approx 1.32$
and $\Omega = 2 E_{\text{so}}$ 
as functions of 
the inverse of the $s$-wave scattering length
and the generalized detuning $\tilde{\delta}$
 using, as in the previous sections, that the 
scattering lengths are the same for all spin channels. 
Using Cs's $a_-^{(1)}$, these parameters
correspond to 
$(k_{\text{so}})^{-1} \approx 10,156a_0$, $E_{\text{so}}/h \approx 0.132$kHz,
and $\Omega/h \approx 0.264$kHz.
Comparison with the $^{87}$Rb experiment at NIST~\cite{spielman},
which uses
$(k_{\text{so}})^{-1} \approx 3,410a_0$ (corresponding to 
$E_{\text{so}}/h \approx 1.786$kHz)
and $\Omega/h$ values ranging from
zero to about $10$kHz, suggests that the parameter
regime covered in 
Fig.~\ref{fig_threebody_energy} is reasonable.
Figure~\ref{fig_threebody_energy}
shows that the three-boson binding energy
for a fixed scattering length is largest for $\tilde{\delta}=0$
(this is where the three-atom
threshold has a degeneracy of six; see Appendix~\ref{appendix_threeatomthreshold}).
In addition, there exists an enhancement of the binding
for $\tilde{\delta}/h \approx 0.301$kHz
(this is where the three-atom
threshold has a degeneracy of four; see Appendix~\ref{appendix_threeatomthreshold}).
As $\tilde{\delta}$ goes to infinity, the trimer
in the presence of the 1D spin-orbit coupling becomes unbound at the same
scattering length
as the corresponding trimer in the absence of spin-orbit coupling
(i.e., at $a_s \approx -20,190a_0$). 

The three-boson binding energy shown in Fig.~\ref{fig_threebody_energy}
is calculated by enforcing that the three-boson threshold has
the same center-of-mass momentum as the trimers (see 
Appendices~\ref{appendix_threeatomthreshold}-\ref{appendix_thresholdall}).
If the detuning $\delta$ is equal to zero, the
generalized detuning $\tilde{\delta}$ 
is directly proportional to the
$z$-component $q_{3,z}$ of the center-of-mass momentum
[see Eq.~(\ref{eq_deltatilde})]. In this case, the
trimer is 
bound maximally 
for $q_{3,z}=0$.
However, for finite detuning $\delta$, the most strongly bound
trimer has a finite center-of-mass momentum.
A similar dependence on the center-of-mass momentum was pointed out 
in Refs.~\cite{pairing_shenoy,han_pu_pra,comment_binding}
for the two-fermion 
system.

The dependence of $\hat{\bar{H}}_{\text{rel}}$
on $q_{N=3,z}$ is a key characteristic
of systems with 1D spin-orbit coupling.
A similar dependence exists for
three-body systems in the presence of 2D or 3D spin-orbit coupling
(in these cases, the relative Hamiltonian depends on two or all three components of
$\vec{q}_{N=3}$) and for
three-body systems on a lattice (in this case,
$\vec{q}_{N=3}$ is a lattice or quasi-momentum vector).
In all works known to us~\cite{cui2014,shi2014, shi2015, nishida}, 
the assumption  $\vec{q}_{N=3}=0$
is made prior to obtaining concrete results.
Table~\ref{tab_literature} contrasts studies for
systems, which possess a center-of-mass momentum dependence,
with the ``standard'' three-boson Efimov 
system  (first row), for which the relative Hamiltonian
is independent of $\vec{q}_{N=3}$.
In the standard Efimov case,
the lowest atom-dimer threshold of the relative Hamiltonian
is given by the energy 
$E_{\text{a}}$ of an atom with vanishing atom momentum vector
$\vec{q}_{\text{a}}$ ($E_{\text{a}}$ is equal to zero)
plus the energy
$E_{\text{d}}(\vec{q}_{\text{d}}=0)$
of a dimer with vanishing dimer
momentum vector $\vec{q}_{\text{d}}$.
When the relative Hamiltonian 
depends on $\vec{q}_{N=3}$, the atom-dimer threshold
needs to be determined carefully, since the
trimer with fixed $\vec{q}_{N=3}$ can break up
  into an atom with finite momentum
  and into a dimer with finite
  momentum in such a way that the generalized 
three-body center-of-mass momentum is conserved.
  Of the many break-up configurations that conserve the
three-body center-of-mass momentum,
  the one with the lowest energy defines the atom-dimer
threshold.
  Table~\ref{tab_literature} shows that the definition of the lowest atom-dimer
  threshold of the
  relative Hamiltonian varies in the literature.
  The definitions employed in Refs.~\cite{cui2014,nishida} disagree with the
  definition used
  in the present work (last row of Table~\ref{tab_literature}). 
While the definition of Ref.~\cite{cui2014}
  may be meaningful in a many-body context
(see also comment~\cite{comment_binding}), we fail
  to see how the definition of Ref.~\cite{nishida} can, in general, 
be correct.

\begin{widetext}

  \begin{table}
  \begin{center}
  \begin{tabular}{c|c|c|c}
    system & $\hat{\bar{H}}_{\text{rel}}=\hat{\bar{H}}_{\text{rel}}(\vec{q}_{N=3})$? & restriction? & atom-dimer threshold (rel. Ham.)\\ \hline
    3-spinless bosons; ``standard'' Efimov scenario~\cite{efimov70} & no & no & $E_{\text{d}}(\vec{q}_{\text{d}}=0)+E_{\text{a}}(\vec{q}_{\text{a}} = 0)$ \\
    FFX; X feels 2D SOC; Borromean binding~\cite{cui2014} & yes & $\vec{q}_{N=3}=0$ & $\mbox{min}_{\vec{q}_{\text{d}}} E_{\text{d}}(\vec{q}_{\text{d}}) + \mbox{min}_{\vec{q}_{\text{a}}} E_{\text{a}}(\vec{q}_{\text{a}})$\\
    FFX; X feels 3D SOC; universal/Efimov trimers~\cite{shi2014,shi2015} & yes & $\vec{q}_{N=3}=0$ & $\mbox{min}_{\vec{q}_{\text{d}}+\vec{q}_{\text{a}}=\vec{q}_{N=3}}
    \left[ E_{\text{d}}(\vec{q}_{\text{d}}) + E_{\text{a}}(\vec{q}_{\text{a}}) \right]$ \\
    BBB quasi-particles on lattice; Efimov trimers~\cite{nishida} & yes & $\vec{q}_{N=3}=0$ & $E_{\text{d}}(\vec{q}_{\text{d}}=0)+E_{\text{a}}(\vec{q}_{\text{a}} = 0)$\\
    BBB; B's feel 1D SOC (this work) & yes & no & $\mbox{min}_{\vec{q}_{\text{d}}+\vec{q}_{\text{a}}=\vec{q}_{N=3}}
    \left[ E_{\text{d}}(\vec{q}_{\text{d}}) + E_{\text{a}}(\vec{q}_{\text{a}}) \right]$
      \end{tabular}
  \caption{Summary of three-particle studies. The standard Efimov scenario (first row)
    is contrasted with three-particle systems for which the relative 
Hamiltonian $\hat{\bar{H}}_{\text{rel}}$ depends
    parametrically on the generalized three-body center-of-mass momentum vector $\vec{q}_{N=3}$.
The last column lists the atom-dimer threshold of the relative Hamiltonian.
    The symbols $E_{\text{d}}$, $E_{\text{a}}$, 
$\vec{q}_{\text{d}}$, and $\vec{q}_{\text{a}}$ denote the energy of the dimer,
    energy of the atom,
    generalized momentum of the dimer, and generalized momentum of the atom,
    respectively. ``SOC'' stands for ``spin-orbit coupling'', ``F'' for ``fermion'',
 ``X'' for a particle different from ``F'', and   ``B'' for ``boson''.
}
\label{tab_literature}
  \end{center}
\end{table}

\end{widetext}

It is proposed that
the center-of-mass momentum dependence can be observed experimentally
by performing atom-loss measurements for fixed
$k_{\text{so}}$, $\Omega$, and $\delta$ on a cold thermal atomic gas.
Tuning the $s$-wave scattering length, one expects---just
as in the case where the spin-orbit coupling is absent---enhanced
losses when the trimer energy is degenerate with the three-atom threshold.
However, in contrast to the standard Efimov scenario, such a degeneracy exists
for a range of scattering lengths provided the trimers embedded in the thermal gas
have different three-body center-of-mass momenta (the exact distribution of center-of-mass
momenta is set by the temperature of the gas sample).
Figure~\ref{fig_threebody_energy} shows that the critical scattering length
$a_-^{(1)}$ changes, for the Cs example, from $-20,190a_0$ for large $\tilde{\delta}$
to $-7,791 a_0$ for $\tilde{\delta}=0$.  
Provided the three-body center-of-mass momenta are spread over the range 
covered on the vertical axis in Fig.~\ref{fig_threebody_energy},
one expects
enhanced losses over the entire
scattering length window.
The difference between the losses in the presence and absence of the spin-orbit coupling terms
can be interpreted as a few-body probe of the breaking of the Galilean invariance in the
presence of spin-orbit coupling.

An important question is whether the 
changes of the loss features related to the lowest state in the second-lowest
manifold will be washed out by finite temperature effects.
A comprehensive answer to this question will require 
performing three-body recombination calculations, which
include thermal averaging, in the presence
of spin-orbit coupling.
Such calculations are beyond the scope of this work.
Given that the energy scales associated with the 
spin-orbit coupling are, for the example considered in Fig.~\ref{fig_threebody_energy},
comparable to $\hbar^2 \kappa_*^2/m$ and 
that the binding energy of the trimer near the three-atom threshold
is much smaller than $\hbar^2 \kappa_*^2/m$, we 
are hopeful that the temperatures 
realized 
in previous Cs experiments ($T \ge 7.7$nK)~\cite{huangPRL} 
are low enough to observe 
the impact of the spin-orbit coupling terms on the loss features.
For example, 
without spin-orbit coupling,
the loss coefficient $L_3$ is maximal around $-20,000a_0$ and
reaches about half of its
maximum value at around $-10,000a_0$ [see Fig.~1(a) of Ref.~\cite{huangPRL}].
In the presence of spin-orbit coupling, the loss feature is expected to be
centered over the range $-20,190a_0$ to $-7,790a_0$, leading to an observable
modification of the shoulder on the less negative scattering length side.
The shape of the shoulder is expected to carry a signature of the non-monotonic
dependence of the critical scattering length $a_-^{(1)}$ on the center-of-mass 
momentum. For example, three different $\tilde{\delta}$
correspond to the same $a_-^{(1)}$ for  $a_-^{(1)} \in [-10,330a_0, -9,160a_0]$
but each $\tilde{\delta}$ 
corresponds to a unique $a_-^{(1)}$ for  
$a_-^{(1)} \in [-20,190a_0, -10,330 a_0]$
and 
$a_-^{(1)} \in [-9,160a_0,-7,790a_0]$.

For the same spin-orbit coupling parameters, 
the critical scattering length $a_-^{(0)}$, associated with the lowest
state in the lowest manifold, displays essentially no dependence on
$\tilde{\delta}$, i.e., the associated three-atom
loss feature should only be minimally
affected by the spin-orbit coupling terms. Intuitively, this can be 
understood by realizing that the energy scales
associated with the spin-orbit coupling parameters 
are much smaller than the binding energy of the lowest-lying 
trimer state. 
The fact that the loss features for the lowest state in the lowest
and second-lowest manifolds are expected to 
be very different can also be understood from the
generalized radial scaling law. Fixing the spin-orbit coupling
parameters corresponds to looking at particular cuts in the
five-dimensional
parameter space as opposed to looking along a specific radial direction.
As a consequence, the loss features for the two manifolds
can be very different even if the scattering lengths at which the loss
features occur are, roughly, spaced by $\lambda_0$.

\section{Concluding remarks}
\label{sec_conclusion}

This work analyzed what happens to the
three-boson Efimov spectrum if 1D spin-orbit
coupling terms, realizable in cold atoms as well
as in photonic crystals and mechanical
setups, are added to the Hamiltonian.
The spin-orbit coupling terms introduce a parametric dependence 
of the relative Hamiltonian on the center-of-mass momentum
vector.
A similar center-of-mass momentum vector dependence
exists for few-body systems with short-range interactions
on a lattice.
The present work mapped out, for the first time, the three-boson
spectrum as a function of the center-of-mass momentum vector.
It was found that the three-boson system in the presence
of 1D spin-orbit coupling obeys a generalized radial scaling law
in a five-dimensional parameter space, which
is associated with
a discrete scaling symmetry. 
Within the framework of effective field theory, the
existence of the discrete scaling symmetry can be rationalized by
scale separation: The discrete scaling symmetry of the standard Efimov scenario
``survives'' provided the additional length scales are much larger than the
ranges of the intrinsic interactions.
While our work focused on 
1D spin-orbit coupling, the discrete scaling symmetry 
should persist for other types of spin-obit coupling schemes as well.

The spin degrees of freedom lead---for the type of
spin-orbit coupling considered in this work---to a 
quadrupling of each Efimov trimer (manifold of four states).
The three-body states in a given manifold
are tied to one of the three two-boson states~\cite{to_be_published}.
The point $(1/a_s,k_{\text{so}},\Omega,\tilde{\delta})=(0,0,0,0)$
serves as an accumulation point for all four states of the manifold, 
i.e., in its vicinity, there exist infinitely many three-body bound states.
The rich structure of two- and three-boson states
should be
amenable to experimental verification.
Due to the dependence of the trimers on the
center-of-mass momentum, the scattering length at which the 
lowest trimer in the second-lowest manifold
merges with the atom-atom-atom threshold
is, in fact, a scattering length window.
Similar scattering length windows exist for the excited states in the
second-lowest manifold.
It was argued that these scattering length windows
should be observable in cold atom loss experiments,
providing a direct few-body signature of the breaking of
the Galilean invariance of systems with
spin-orbit coupling.

If one considers a cut in the generalized five-dimensional
Efimov plot, energy levels are not spaced by the scaling factor
$(\lambda_0)^2$.
Let us consider the situation where 
$a_s$ is infinitely large and 
where $k_{\text{so}}$, $\Omega$, and $\tilde{\delta}$
are finite.
In this case, the low-energy scales associated with the 1D spin-orbit
coupling terms lead
to a cut-off of the hyperradial
$-1/R^2$ Efimov potential curve ($R$ denotes the three-body hyperradius).
As a consequence, the number of three-body bound states
at unitarity is not infinitely large.
Albeit due to a different mechanism,
this is similar in spirit
to the disappearance of Efimov states if an Efimov
trimer is placed into a gas of bosons or
fermions~\cite{zinner2013,zhou2011}.
This is also similar in spirit to a rather different system, namely
the H$^-$ ion.
Taking only Coulomb interactions into account,
one obtains a $-1/r^2$ attraction~\cite{gailitis},
where $r$ is the distance between the extra electron and the atom.
Relativistic effects
introduce an additional length scale, which renders the number of bound
states finite~\cite{gailitis}.

The calculations in this work were performed 
assuming that the interactions between the
different spin-channels are all equal.
If one of the scattering lengths is large and tunable
while the others are close to zero,
the discrete scaling symmetry should still hold (approximately).
To find the functional form of the energies for this scenario, the 
spectrum has to be recalculated. 

The study presented should be viewed as a first step toward
uncovering the rich three- and higher-body physics
that emerges as a consequence of the unique coupling between the
relative and center-of-mass degrees of freedom
in cold atom systems in the presence of artificial Gauge fields.
While somewhat different in nature, the coupling of these degrees of freedom
in the relativistic Klein Gordon and Dirac equations and quantum
field theories has captured
physicists' imagination for many decades.

\section{Acknowledgement}
\label{acknowledgement}
Support by the National Science Foundation through
grant numbers  
PHY-1509892 and PHY-1745142
is gratefully acknowledged.
This work used the Extreme Science and Engineering
Discovery Environment (XSEDE), which is supported by
NSF Grant No. OCI-1053575.
Some of the computing for this project was performed at the OU 
Supercomputing Center for Education and Research (OSCER) at the University of Oklahoma (OU).

\appendix
\section{Jacobi coordinates}
\label{appendix_basics}
The single-particle and Jacobi coordinates employed in this 
work are related through the matrix $\underline{U}$~\cite{ECGbook},
\begin{eqnarray}
(\vec{\rho}_1,\cdots,\vec{\rho}_N)^T
=\underline{U} (\vec{r}_1,\cdots,\vec{r}_N)^T,
\end{eqnarray}
where 
\begin{eqnarray}
\underline{U} =
\left(
\begin{array}{cc}
1 & -1 \\
1/2 & 1/2
\end{array}
\right)
\end{eqnarray}
for the equal-mass two-particle system and
\begin{eqnarray}
\underline{U} =
\left(
\begin{array}{ccc}
1 & -1 & 0 \\
1/2 & 1/2 & -1 \\
1/3 & 1/3 & 1/3
\end{array}
\right)
\end{eqnarray}
for the equal-mass three-particle system.
The transformation matrix $\underline{U}$
also defines the matrices $\hat{\Sigma}_{j,z}$
($j=1,\cdots,N-1$). 
For the two-body system,
we have
\begin{eqnarray}
\hat{\Sigma}_{1,z}=
\hat{\sigma}_{1,z} \otimes I_2 -
I_1 \otimes \hat{\sigma}_{2,z} . 
\end{eqnarray}
For the three-body system, 
we have
\begin{eqnarray}
\hat{\Sigma}_{1,z}=
\hat{\sigma}_{1,z} \otimes I_2 \otimes I_3 -
I_1 \otimes \hat{\sigma}_{2,z} \otimes I_3
\end{eqnarray}
and
\begin{eqnarray}
\hat{\Sigma}_{2,z}&=&
\frac{1}{2}
\left(
\hat{\sigma}_{1,z} \otimes I_{2} \otimes I_3 +
I_1 \otimes \hat{\sigma}_{2,z} \otimes I_3
\right)
- \nonumber \\
&& I_1 \otimes I_2 \otimes \hat{\sigma}_{3,z} . 
\end{eqnarray}

\section{Explicitly correlated basis set expansion approach}
\label{appendix_ecg}
This appendix discusses our approach to solving the few-particle time-independent
Schr\"odinger equation using a basis set expansion in terms of
explicitly correlated Gaussian basis functions,
which contain
non-linear variational parameters
that are optimized semi-stochastically.

As discussed in the main text,
we are interested in bound states, i.e.,
eigen states that approach zero at large interparticle distances.
To solve the time-independent Schr\"odinger equation,
we expand the relative portion
$\Phi_{\text{rel}}$ of the eigen state $\Psi$ sought 
in terms of a set of non-orthogonal
eigen functions $\psi_j$~\cite{ECGbook,ECGrmp},
\begin{eqnarray}
\Phi_{\text{rel}} = \sum_{j=1}^{N_b} c_j \psi_j .
\end{eqnarray}
The basis functions $\psi_j$ depend on the relative spatial and the spin
degrees of freedom,
\begin{eqnarray}
\label{eq_basisfunction1}
  \psi_j= {\hat{\cal{S}}}
  \left( \phi_j(\vec{\rho}_1,\cdots,\vec{\rho}_{N-1}) \chi_j
  \right),
\end{eqnarray}
where $\chi_j$ denotes an $N$-particle spin function
that is chosen from the complete set of 
$2^N$ possible spin functions.
The spatial parts $\phi_j$ are written in terms of 
a total of $(N-1)(N/2+3)$
non-linear variational parameters
$d_{kl}^{(j)}$ and $\vec{s}_k^{(j)}$,
\begin{eqnarray}
\label{eq_basisfunction}
\phi_j = \exp \left( - \sum_{k<l}^N \frac{r_{kl}^2}{2 d_{kl}^{(j)}} +
\sum_{k=1}^{N-1} \imath \vec{s}_k^{(j)} \cdot \vec{\rho}_k \right) .
\end{eqnarray}
Here, the superscript ``$(j)$'' serves to remind us that each basis
function is characterized by a set of non-linear variational
parameters.
The non-linear variational
parameters $d_{kl}^{(j)}$ 
determine the widths of the Gaussian factors
of the basis functions.
These widths are governed, roughly, by the two-body interaction terms in
the Hamiltonian.
The
non-linear parameters $\vec{s}_k^{(j)}$ determine
the spatial oscillations due to the $k_{\text{so}}$-dependent one-body terms.  
We do find that the values of $\vec{s}_k^{(j)}$ 
can depend quite strongly on the spin basis function
considered. This shows that the parameters $\vec{s}_k^{(j)}$
govern, to leading order, the interplay between the spatial and
spin degrees of freedom.
Of course, strictly speaking, 
the influence of the single-particle and two-particle
interaction terms in the Hamiltonian on the eigen states cannot 
be separated; rather, the eigen states are the result of the relative
importance of each of these terms 
and the kinetic energy terms.
The $N(N-1)/2$ non-linear parameters $d_{kl}^{(j)}$ 
and the $3(N-1)$ non-linear parameters 
contained in $\vec{s}_k^{(j)}$ are
optimized
semi-stochastically, i.e., the basis set is constructed so as to
minimize the energy of the eigen state under study.
The symmetrizer $\hat{\cal{S}}$ in Eq.~(\ref{eq_basisfunction1})
ensures that the basis functions, and hence the eigen state sought,
are fully symmetrized. For the two-boson system, e.g.,
the symmetrizer reads $\hat{\cal{S}}=(1+\hat{P}_{12})/\sqrt{2}$,
where $\hat{P}_{12}$ exchanges the spatial and spin degrees of
freedom of particles
1 and 2. For $N$ identical particles, the symmetrizer contains $N!$ terms.

The linear parameters $c_j$ are determined by
solving the generalized eigen value problem that depends
on the Hamiltonian matrix and, due to the non-orthogonality
of the basis functions, the overlap matrix~\cite{ECGbook}.
The results presented in this paper utilize
basis sets consisting of up to $N_b = 1,800$ basis 
functions.
Key strengths of the numerical approach employed are that
the Hamiltonian and overlap matrix elements 
have compact analytical expressions and that the 
non-linear variational parameters can be adjusted so
as to capture correlations that occur on scales
smaller than $r_0$  and larger than $(k_{\text{so}})^{-1}$.

The energies for the $N=2$ system depend on the parameters
$a_s$, $k_{\text{so}}$, $\Omega$, and $\tilde{\delta}$.
For $N=3$, they additionally depend on $\kappa_*$.
In practice,
we set $\vec{q}_N=0$ and scan
$\Omega/E_{\text{so}}$, $\delta/E_{\text{so}}$, and $a_s k_{\text{so}}$
(for $N=3$, we fix $\kappa_*$).
To obtain the relative
eigen energies $E$ and eigen states $\Phi_{\text{rel}}$
for finite ${q}_{N,z}$,
we do not need to redo the numerical calculations.
Instead, we use the ``conversion'' implied by 
Eq.~(\ref{eq_deltatilde}), i.e., the relative eigen
energy and eigen states are obtained by 
changing from $\delta$ to $\tilde{\delta}$.
The full
eigen states $\Psi$
are obtained by multiplying the $\vec{q}_N=0$
eigen states by the center-of-mass 
piece $\Phi_{\text{cm}}$.

As in the case without spin-orbit coupling~\cite{rakshit_daily_blume}, it
tends to be more efficient to describe each relative
eigen state $\Phi_{\text{rel}}$
by its own basis set as opposed to constructing one basis set
that describes multiple eigen states well. 
We refer to the eigen state sought---this 
could be the ground state or one
of the excited states---as target state.
To construct the basis, we add one basis function
at a time. Let us assume that we 
have a basis set of size $N_{\text{initial}}$ and that we want to 
enlarge the basis set by one basis function.
To do this, we generate 
$N_{\text{trial}}$
trial basis functions 
($N_{\text{trial}}$ is of the order of a few hundred
to a few thousand) 
and calculate
the energy $E_k$ ($k=1,\cdots,N_{\text{trial}}$) 
of the target state for each of these enlarged basis sets.
Assuming that
the generalized eigen value problem for 
the basis set of size $N_{\text{initial}}$ has been solved, the
trial energies can be calculated via a root-finding 
procedure~\cite{ECGbook},
which is, generally, computationally 
significantly faster 
than solving the
generalized eigen value problem.
The basis function to be added is determined by which
of the $N_{\text{trial}}$ trial energies is lowest, i.e., 
by looking for the trial basis function that lowers the energy
of the target state the most.
After the ``best'' trial function has been added, the generalized 
eigen value problem is solved for the basis set of size
$N_{\text{initial}}+1$ and the procedure is repeated to generate a basis set
of size $N_{\text{initial}}+2$.

The convergence of the energy 
with increasing basis set size can be improved significantly
by choosing ``good'' basis functions, i.e., by
generating basis functions that efficiently cover the entire
Hilbert space.
Conversely, if the basis set is not constructed carefully,
the energy may not even converge. 
In our implementation, the parameters
$d_{kl}^{(j)}$ and $\vec{s}_k^{(j)}$ that characterize the
spatial part of the
basis functions are choosen from carefully adjusted
parameter windows. For example, the $d_{kl}^{(j)}$
are chosen so as to cover length scales 
ranging from less than 
$r_0$ to a few times the length set by the
binding energy of the target state. Since the target
energy is, in general, not known {\em{a priori}}, 
the parameter windows are typically refined based
on results obtained in preliminary calculations.
To choose parameter windows 
for the $x$-, $y$- and $z$-components of 
${\vec{s}}_k^{(j)}$, we are 
guided by the non-interacting two- and three-particle
dispersion curves. For example,
if the non-interacting two-particle
dispersion along the $z$-coordinate exhibits
two minima,
we choose
${s}_{1,z}^{(j)}$ uniformly from the windows
$[-s_{\text{max},z},-s_{\text{min},z}]$ and $[s_{\text{min},z},s_{\text{max},z}]$,
with the windows including the momenta at which
the dispersion curve is minimal.
The parameters ${s}_{1,x}^{(j)}$ and ${s}_{1,y}^{(j)}$
are selected from windows that include zero.
The widths of the parameter windows are adjusted through
an educated trial and error procedure.
Among other things, we check if the
parameters selected by the code are clumped in a particular region of the parameter
space. 

\section{Three-atom threshold}
\label{appendix_threeatomthreshold}
The three-atom system with center-of-mass momentum $\vec{q}_N$,
$N=3$, 
is bound if its energy is lower than that
of three infinitely far separated atoms with the same center-of-mass 
momentum
and, if a two-body bound state exists, lower than that of
an infinitely far separated dimer and atom with the same center-of-mass
momentum.
To determine the 
lowest three-body scattering threshold,
one thus needs 
to know the lowest dimer binding energy 
for all two-body
center-of-mass momenta.
As a consequence, the three-body scattering threshold
depends on $\tilde{\delta}/E_{\text{so}}$ and $\Omega/E_{\text{so}}$ 
as well as on $a_s k_{\text{so}}$.
We define the scattering threshold
$E_{\text{th}}$ using $\hat{\bar{H}}_{\text{rel}}$
and denote the eigen energies of 
$\hat{\bar{H}}_{\text{rel}}$
by $E$.

\begin{figure}
\vspace*{0cm}
\hspace*{-0.5cm}
\includegraphics[width=0.5\textwidth]{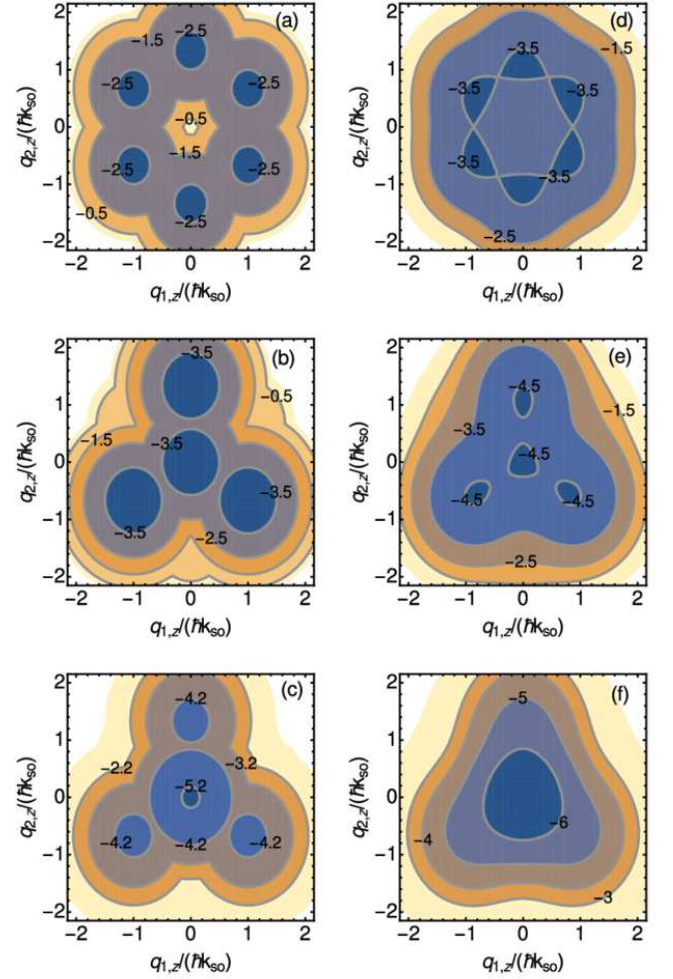}
\caption{(Color online)
Lowest 
non-interacting relative 
three-atom dispersion
curve.
The contours show the lowest 
non-interacting relative three-atom
dispersion curve, in units of $E_{\text{so}}$,
as functions of 
$q_{1,z}$ and $q_{2,z}$.
Panels~(a)-(c) are for
$\Omega/E_{\text{so}}=0$ and $\tilde{\delta}/E_{\text{so}}=0$,
$8/3$, and $3.5$, respectively.
Panels~(d)-(f) are for
$\Omega/E_{\text{so}}=2$ and $\tilde{\delta}/E_{\text{so}}=0$,
$2.278$, and $4$, respectively.
}
\label{fig_3b_dispersion}
\end{figure}

We start by determining the lowest relative three-body
scattering threshold in the absence of two-body bound states.
In this case, 
the lowest relative
scattering threshold is
determined by 
the minimum energy of the 
non-interacting relative 
dispersion curves for fixed $\tilde{\delta}/E_{\text{so}}$ and $\Omega/E_{\text{so}}$.
The dispersion curves depend on
two relative momenta 
(namely $q_{1,z}$ and $q_{2,z}$)
and the total number of dispersion curves
is eight.
Figures~\ref{fig_3b_dispersion}(a)-\ref{fig_3b_dispersion}(c)
are for the uncoupled case ($\Omega=0$) and
$\tilde{\delta}/E_{\text{so}}=0$, $8/3$, and
$3.5$, respectively.
The number of global minima changes from six
for
$\tilde{\delta}=0$ [see Fig.~\ref{fig_3b_dispersion}(a)]
to three for
$0<\tilde{\delta}/E_{\text{so}}< 8/3$ (not shown)
to four for $\tilde{\delta}/E_{\text{so}}= 8/3$
[see Fig.~\ref{fig_3b_dispersion}(b)]
to one for $\tilde{\delta}/E_{\text{so}}> 8/3$
[see Fig.~\ref{fig_3b_dispersion}(c)].
A finite Raman coupling strength $\Omega$
introduces a coupling between the different spin channels.
As an example,
Figs.~\ref{fig_3b_dispersion}(d)-\ref{fig_3b_dispersion}(f)
show the lowest relative non-interacting dispersion curves
for
$\Omega/E_{\text{so}}=2$ and
$\tilde{\delta}/E_{\text{so}}=0$, $2.278$, and
$4$, respectively.
As for vanishing Raman coupling, the number of global
minima changes from six to three (not shown)
to four to one
with increasing $\tilde{\delta}$.
However, the critical generalized detuning $\tilde{\delta}$
at which these changes occur differs for $\Omega/ E_{\text{so}}=2$ and 
$\Omega=0$.

\begin{figure}
\vspace*{0.cm}
\hspace*{-0.5cm}
\includegraphics[width=0.32\textwidth]{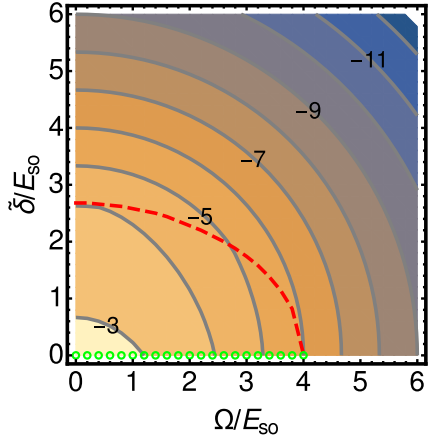}
\caption{(Color online)
Lowest relative three-atom scattering threshold.
The contours
show the energy $E_{\text{th}}^{\text{aaa}}$, in units of $E_{\text{so}}$,
of the lowest three-atom scattering threshold
as functions of $\Omega/E_{\text{so}}$ and $\tilde{\delta}/E_{\text{so}}$.
The thick open circles and thick dashed line indicate the
parameter combinations $(\tilde{\delta}/E_{\text{so}},\Omega/E_{\text{so}})$
at which the degeneracy of the lowest three-atom scattering threshold
is six and four, respectively.
}
\label{fig_3b_aaathreshold}
\end{figure}

The minimum of the non-interacting relative
three-atom dispersion curves defines, assuming two-body bound states
are absent, the lowest three-atom scattering threshold.
Figure~\ref{fig_3b_aaathreshold}
shows the lowest three-atom scattering threshold energy
$E^{\text{aaa}}_{\text{th}}$
as functions of $\tilde{\delta}/E_{\text{so}}$ and $\Omega/E_{\text{so}}$.
The thick open circles and thick dashed line indicate
the parameter combinations at which the
number of global minima is six and four,
respectively.
For parameter combinations above the thick open circles
and below the thick dashed line the number of global
minima of the lowest
non-interacting relative dispersion curve
is equal to three. Above the thick dashed line the
number of global minima is equal to one.
If we assume
that the three-body
binding energy is, approximately, largest when the degeneracy of the lowest
non-interacting
relative
 dispersion curve
is largest, then Fig.~\ref{fig_3b_aaathreshold}
suggests that the three-body system on the
negative scattering length side, provided two-body bound states
are absent, is enhanced the most compared
to the energy of the system without spin-orbit
coupling when $\tilde{\delta}=0$
or $q_{3,z}=-3m \delta / (2 \hbar k_{\text{so}})$.
The main text shows that this reasoning provides an intuitive
understanding for the behavior of the lowest three-boson
state in each manifold.
However, 
the situation for the excited states in a manifold is more 
intricate~\cite{to_be_published}.

\section{Atom-dimer threshold}
\label{appendix_atomdimerthreshold}

As already mentioned, the determination of the
lowest atom-dimer scattering threshold requires knowledge of the
dimer binding energy and the single-particle dispersion curve.
Since the $z$-component 
of the center-of-mass momentum $q_{1,z}$ of the dimer,
formed by particles 1 and 2, can be written as a linear
combination of $q_{2,z}$ and $q_{3,z}$, $q_{1,z}$ is not a free parameter.
As a consequence, the atom-dimer dispersion curves
depend only on $q_{2,z}$ but not on $q_{1,z}$.
Physically, this makes sense since the three-body system
breaks up into two units (a dimer and an atom), with the
momentum between the two units determining the division
of the three-body momentum among the dimer and the atom.

To quantify this, we 
rewrite the Hamiltonian $\hat{\bar{H}}_{\text{rel}}$
by arbitrarily singling out the third atom
and treating the expectation value $q_{2,z}$ of
$\hat{q}_{2,z}$ as a parameter,
\begin{eqnarray}
\label{eq_ham_ad}
\hat{\bar{H}}_{\text{rel}}
\big|_{\langle \hat{q}_{2,z} \rangle =q_{2,z}} &=&
\hat{H}_{12}(q_{2,z}) \otimes I_3 + \nonumber \\
&& I_1 \otimes I_2 \otimes \hat{H}_3(q_{2,z})
+\nonumber \\
&&V_{\text{coupling}}.
\end{eqnarray}
Here, the ``dimer Hamiltonian'' $\hat{H}_{12}(q_{2,z})$ 
reads
\begin{eqnarray}
\hat{H}_{12}(q_{2,z})&=&
\left( \frac{\hat{\vec{q}}_1^2}{2 \mu_1} + {V}_{\text{2b}}(r_{12})
\right) I_1 \otimes I_2 
+ \nonumber \\
&&\frac{\hbar k_{\text{so}} q_{1,z}}{m} \left(
\hat{\sigma}_{1,z} \otimes I_2 - I_1 \otimes \hat{\sigma}_{2,z}
\right)
+ \nonumber \\
&& \left(
\frac{\hbar k_{\text{so}} q_{2,z}}{2m} + \frac{\tilde{\delta}}{2}
\right) 
\left(
\hat{\sigma}_{1,z} \otimes I_2 + I_1 \otimes \hat{\sigma}_{2,z} 
\right) 
+ \nonumber \\
&& \frac{\Omega}{2}
\left(
\hat{\sigma}_{1,x} \otimes I_2 + I_1 \otimes \hat{\sigma}_{2,x}
\right).
\end{eqnarray}
Identifying $\tilde{\delta}_{12,\text{eff}}$,
\begin{eqnarray}
\label{eq_deltatilde_dimereff}
\frac{\tilde{\delta}_{12,\text{eff}} }{2}
=
\frac{\hbar k_{\text{so}} q_{2,z}}{2m} + \frac{\tilde{\delta}}{2},
\end{eqnarray}
as a new effective dimer detuning,
the eigen energies of $\hat{H}_{12}(q_{2,z})$
are the same as those of the two-body
Hamiltonian.
The ``atom Hamiltonian'' $\hat{H}_3(q_{2,z})$,
\begin{eqnarray}
\hat{H}_{3}(q_{2,z})= \nonumber \\
\frac{{\vec{q}}_2^2}{2 \mu_2} \otimes I_3 + 
\left(
-\frac{\hbar k_{\text{so}} {q}_{2,z}}{m} + \frac{\tilde{\delta}}{2}
\right) \hat{\sigma}_{3,z} + \frac{\Omega}{2} \hat{\sigma}_{3,x},
\end{eqnarray}
describes the Jacobi particle with mass $\mu_2$
and effective atom detuning $\tilde{\delta}_{3,\text{eff}}$,
where
\begin{eqnarray}
\label{eq_deltatilde_atomeff}
\frac{\tilde{\delta}_{3,\text{eff}}}{2} =
-\frac{\hbar k_{\text{so}} {q}_{2,z}}{m} + \frac{\tilde{\delta}}{2}.
\end{eqnarray}
Note that the effective dimer detuning
$\tilde{\delta}_{12,\text{eff}}$ and the
effective atom detuning $\tilde{\delta}_{3,\text{eff}}$
depend on the ``true detuning'' $\delta$, which is 
fixed by the experimental
set-up,
on the $z$-component $q_{3,z}$ of the three-body center-of-mass 
momentum,
which is a conserved quantity,
and on $q_{2,z}$, which is treated as a parameter.
Assuming that the distance between the
center-of-mass of the dimer and the atom
is large compared to the size of the dimer and compared to 
the ranges $r_0$ and $R_0$ of the two-
and three-body interactions,
the coupling term $V_{\text{coupling}}$,
\begin{align}
\label{eq_coupling_ad}
  V_{\text{coupling}}= 
  [V_{\text{2b}}(r_{13}) + V_{\text{2b}}(r_{23}) + V_{\text{3b}}(r_{123})] I_1 \otimes I_2 \otimes I_3,
\end{align}
can be set to zero.
Thus, the $q_{2,z}$-dependent relative
atom-dimer dispersion curves are obtained
by adding the eigen energies
of $\hat{H}_{12}$ and $\hat{H}_3$, which depend parametrically on $q_{2,z}$.

Equations~(\ref{eq_ham_ad})-(\ref{eq_coupling_ad})
assumed that the dimer is formed by atoms 1 and 2. Alternatively, the
dimer could be formed by atoms 1 and 3 or by atoms 2 and 3.
These alternative divisions yield atom-dimer dispersion curves
that depend on the $z$-component of the 
momentum that is associated with the distance
vector between particle 2 and the center-of-mass of the 13-dimer and
the $z$-component of the momentum that is associated with the distance
vector between particle 1 and the center-of-mass of the 23-dimer, respectively.
Since we are considering three identical bosons, the three divisions are 
equivalent.
In what follows, we use $q_{\text{ad},z}$ to reflect that we
could single out any of the three atoms.
The corresponding atom-dimer energy is denoted by $E^{\text{ad}}_{\text{th}}$.

\begin{figure}
\vspace*{0.5cm}
\hspace*{0cm}
\includegraphics[width=0.4\textwidth]{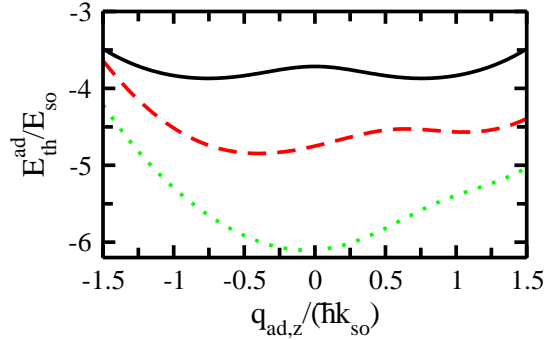}
\caption{(Color online)
Relative atom-dimer dispersion curves
for $(a_s k_{\text{so}})^{-1} = 0.01128$ and $\Omega/E_{\text{so}}=2$.
The solid, dashed, and dotted lines show the energy $E^{\text{ad}}_{\text{th}}$
as a function of the 
$z$-component
$q_{\text{ad},z}$ of the atom-dimer
momentum 
for $\tilde{\delta}/E_{\text{so}} =0, 2.287$, and $3.5$, respectively.
}
\label{fig_atomdimerdispersion}
\end{figure}

Since there exist up to three two-boson bound states~\cite{to_be_published},
the three-boson system supports up to six atom-dimer dispersion curves 
(there could be four or two).
As an example,
Fig.~\ref{fig_atomdimerdispersion} shows the energy $E^{\text{ad}}_{\text{th}}$ 
of the lowest relative
atom-dimer dispersion curve as a function of $q_{\text{ad},z}$
for $(a_s k_{\text{so}})^{-1}=0.01128$,
$\Omega/E_{\text{so}}=2$,
and various $\tilde{\delta}$,
i.e., $\tilde{\delta}=0, 2.287$, and $3.5$.
The system supports,
for this Raman coupling strength
and scattering length, one weakly-bound two-boson state 
for all two-body center-of-mass momenta.
For $\tilde{\delta}=0$ (solid line in Fig.~\ref{fig_atomdimerdispersion}),
the atom-dimer dispersion is symmetric with respect to $q_{\text{ad},z}=0$
and supports two global minima at finite $q_{\text{ad},z}$.
The break-up into a dimer and an atom is energetically most
favorable when $q_{\text{ad},z}/(\hbar k_{\text{so}})$ is equal to
$\pm 0.76$. 
This translates, 
using Eqs.~(\ref{eq_deltatilde_dimereff}) and (\ref{eq_deltatilde_atomeff}), into 
$\tilde{\delta}_{12,\text{eff}}/E_{\text{so}}= \pm 1.52$ and 
$\tilde{\delta}_{3,\text{eff}}/E_{\text{so}}= \pm 3.04$.
For $\tilde{\delta}>0$ (the dashed and dotted lines 
in Fig.~\ref{fig_atomdimerdispersion} are for 
$\tilde{\delta}/E_{\text{so}}=2.287$ and $3.5$, respectively),
the atom-dimer dispersions are asymmetric with respect to $q_{\text{ad},z}=0$
and exhibit a global minimum at negative $q_{\text{ad},z}$, which approaches
$q_{\text{ad},z}=0$ in the $\tilde{\delta} \rightarrow \infty$ limit.
Intuitively, the asymmetry can be understood by realizing that the
atom and the dimer already see a detuning. Thus, moving in the
positive momentum direction is not equivalent to moving
in the negative momentum direction.
The minimum of the lowest 
relative atom-dimer dispersion curve decreases with
increasing $\tilde{\delta}$.

\section{Three-body threshold}
\label{appendix_thresholdall}
\begin{figure}
\vspace*{0.0cm}
\hspace*{-0.5cm}
\includegraphics[width=0.32\textwidth]{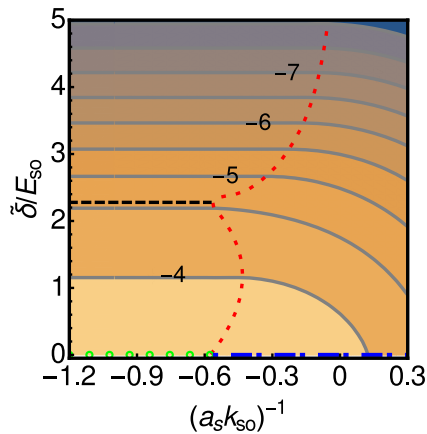}
\caption{(Color online)
Lowest relative three-boson scattering threshold for $\Omega/E_{\text{so}}=2$.
The contours show
the energy $E_{\text{th}}$, in units of $E_{\text{so}}$,
of the lowest three-boson scattering threshold
as functions of $(a_s k_{\text{so}})^{-1}$ and $\tilde{\delta}/E_{\text{so}}$.
To the left of the thick dotted line, bound dimers are not supported
and the lowest
threshold is given by $E_{\text{th}}^{\text{aaa}}$.
In this regime, the lowest scattering threshold is
independent of $(a_s k_{\text{so}})^{-1}$.
To the right of the thick dotted line, a weakly-bound bosonic dimer exists
and the lowest threshold is given by
$E_{\text{th}}^{\text{ad}}$.
In this regime, the
lowest scattering threshold depends on $(a_s k_{\text{so}})^{-1}$.
The thick open circles and thick dashed line
mark the $((a_s k_{\text{so}})^{-1},\tilde{\delta}/E_{\text{so}})$
combinations for which the lowest three-atom threshold has
a degeneracy of six and four, respectively.
The thick dash-dotted line marks the $((a_s k_{\text{so}})^{-1},\tilde{\delta}/E_{\text{so}})$
combinations for which the lowest atom-dimer threshold has
a degeneracy of two.
In the region encircled by
the thick open circles, the thick dotted line, the thick dashed line,
and the left edge of the figure
the degeneracy of the three-atom threshold is
equal to three.
In the region encircled by
the thick dashed line, the thick dotted line, the upper edge of the figure,
and the left edge of the figure,
the degeneracy of the three-atom threshold is
equal to one.
In the region encircled by
the thick dash-dotted line, the right edge of the figure,
the upper edge of the figure, and the thick dotted line
the degeneracy of the atom-dimer threshold is
equal to one.
The range of scattering lengths shown on the horizontal axis is, even though
different units are used, the same as in Fig.~\ref{fig_threebody_energy}.
}
\label{fig_3b_threshold}
\end{figure}

The three-boson threshold is given by the
minimum of the lowest three-atom threshold
and the lowest atom-dimer threshold.
It depends on the values of $\Omega$, $k_{\text{so}}$, $\tilde{\delta}$,
and the $s$-wave scattering length. Using $k_{\text{so}}$ and $E_{\text{so}}$
as units,
Fig.~\ref{fig_3b_threshold} shows a contour plot of the lowest
relative three-boson threshold as functions of
the generalized detuning $\tilde{\delta}/E_{\text{so}}$ and the inverse
$(a_s k_{\text{so}})^{-1}$ of the $s$-wave scattering length for $\Omega/E_{\text{so}}=2$.
As already discussed, the parameter regime in which
two-boson bound states exist
depends on the value of $a_s$.
Correspondingly, the thick dotted line, which marks the
separation of the region in which
the  three-atom threshold has the lowest energy (to
the left of the thick dotted line)
and that in which the atom-dimer threshold has the lowest energy
(to the right of the thick dotted line), shows a distinct dependence on
the $s$-wave scattering length.
For large $\tilde{\delta}$, the thick dotted line approaches
the $(a_s k_{\text{so}})^{-1}=0$ line.
For a fixed $\tilde{\delta}$, the energy $E^{\text{ad}}_{\text{th}}$
of the lowest atom-dimer threshold decreases with
increasing $(a_s k_{\text{so}})^{-1}$. This can be traced back to the
increase of the binding energy
of the two-boson ground state with increasing $(a_s k_{\text{so}})^{-1}$.
The parameter combinations with the largest degeneracy of the
scattering threshold are shown
by the thick open circles (three-atom threshold; the degeneracy
is six) and the
thick dash-dotted line (atom-dimer threshold; the degeneracy is two).
For all $a_s$, the largest degeneracy
of the scattering threshold is found for $\tilde{\delta}=0$.
As discussed in the main text, 
our numerical three-boson calculations show that the binding energy
of the most strongly-bound state in each manifold,
determined as functions of $(a_s k_{\text{so}})^{-1}$ and
$\tilde{\delta}/E_{\text{so}}$, is
largest 
for vanishing $\tilde{\delta}$, i.e., where the degeneracy of the lowest three-boson
scattering threshold is maximal.

\begin{thebibliography}{100}
\bibitem{footnote2}
The normalization factor of the plane wave center-of-mass wave function is not written out for 
notational convenience.
\bibitem{comment_binding}
When we apply our formalism to the two-identical
fermion system, our binding energy 
agrees with the results obtained by Shenoy~\cite{pairing_shenoy} but
disagrees with the results presented in Ref.~\cite{han_pu_pra}.
Reference~\cite{han_pu_pra} calculated the two-fermion binding energy by
allowing the $z$-component of the center-of-mass
momentum of the
eigen state of the interacting system
and that of the two-atom state associated
with the threshold with respect to which the binding energy
is being measured to be different.
The definition employed
[see Eq.~(14) of Ref.~\cite{han_pu_pra}] implies that the
$z$-component of the center-of-mass momentum is not a conserved quantity.
A straightforward calculation, however,
establishes that the
three components of the center-of-mass momentum are individually
conserved (see, e.g., Ref.~\cite{qingze_thesis}).
Correspondingly,
the quantity considered
in Ref.~\cite{han_pu_pra}
is inconsistent with the accepted definition of the
term ``two-body binding energy''.
The quantity defined in Ref.~\cite{han_pu_pra}
does, however, become meaningful in a many-body context.
To see this, we consider a two-body bound state
with total energy $E_2(\vec{q}_2)=E+ \vec{q}_2^2/(4m)$
embedded into a many-body
background (here, $E$ is the eigen energy of the relative
two-particle Hamiltonian; $E$ depends on $q_{2,z}$).
Assuming that the dimer can reduce its energy
via center-of-mass changing
inelastic collisions with the background atoms,
the lowest energy of the dimer is given by
$\min_{\vec{q}_2} E_2(\vec{q}_2)$.
The dimer can then be considered stable
if the global
energy $E_{\text{global},2}$,
$E_{\text{global},2}=
\min_{\vec{q}_2} E_2(\vec{q}_2)- \min_{\vec{q}_2} 
[E_{\text{th}} + \vec{q}_2^2/(4m)]$
($E_{\text{th}}$ denotes the lowest relative atom-atom scattering threshold
for fixed $q_{2,z}$)
is negative.
For the global energy
$E_{\text{global},2}$, referred to as binding energy in Ref.~\cite{han_pu_pra},
to be meaningful,
the dimer needs to be embedded into a background that can, via inelastic
collisions, absorb energy on a time scale that is fast
compared to the time scales associated with other loss processes.
\end{thebibliography}
\end{document}